\documentclass[twocolumn,reprint,prb,amsmath,amssymb,aps,superscriptaddress,floatfix]{revtex4-2}

\usepackage{xr-hyper}
\usepackage{hyperref}
\usepackage{comment}
\hypersetup{colorlinks,urlcolor=blue,citecolor=blue}
\usepackage{graphicx}
\usepackage[percent]{overpic}
\usepackage{dcolumn}
\usepackage{bm}
\usepackage{lipsum} 
\usepackage{ulem} 
\usepackage[dvipsnames]{xcolor}
\usepackage{capt-of}
\usepackage{float} 

\begin{document}

\title{Tuning charge density wave order and structure via uniaxial stress in a stripe-ordered cuprate superconductor}

\author{Naman~K.~Gupta}
\affiliation{Department of Physics and Astronomy, University of Waterloo, Waterloo, Ontario N2L 3G1, Canada}

\author{Ronny~Sutarto}
\affiliation{Canadian Light Source, University of Saskatchewan, Saskatoon, Saskatchewan, S7N 2V3, Canada}

\author{Rantong~Gong}
\affiliation{Department of Physics and Astronomy, University of Waterloo, Waterloo, Ontario N2L 3G1, Canada}

\author{Stefan~Idziak}
\affiliation{Department of Physics and Astronomy, University of Waterloo, Waterloo, Ontario N2L 3G1, Canada}

\author{Hiruy~Hale}
\affiliation{Department of Physics and Astronomy, University of Waterloo, Waterloo, Ontario N2L 3G1, Canada}

\author{Young-June~Kim}
\affiliation{Department of Physics, University of Toronto, Toronto, M5S 1A7, Canada}

\author{David~G. Hawthorn}
\thanks{Corresponding author: david.hawthorn@uwaterloo.ca}
\affiliation{Department of Physics and Astronomy, University of Waterloo, Waterloo, Ontario N2L 3G1, Canada}

\begin{abstract}
Spin and charge density wave order in the cuprates are known to compete with superconductivity. In the stripe order (La,M)$_2$CuO$_4$ family of cuprates, spin and charge order occur as unidirectional order that can be stabilized by symmetry breaking structural distortions, such as the low temperature tetragonal (LTT) phase. Here we examine the interplay between structure and the formation of charge density wave (CDW) order in the LTT phase of La$_{1.475}$Nd$_{0.4}$Sr$_{0.125}$CuO$_4$ by applying uniaxial stress to distort the structure and influence the formation of CDW order. Using resonant soft x-ray scattering to measure both the CDW order and (0~0~1) structural-nematic Bragg peaks, we find that the application of uniaxial stress along the Cu-O bond direction suppresses the (0~0~1) peak and has the net effect of reducing CDW order, but does so only for CDW order propagating parallel to the applied stress. We connect these observations to previous work showing an enhanced superconducting transition temperature under uniaxial stress; providing insight into how CDW, superconductivity, nematicity, and structure are related and can be tuned relative to one another in cuprates. 
\end{abstract}
\maketitle

In the cuprate superconductors, spin and charge density wave orders~\cite{tranquada1995evidence, Ghiringhelli12, Chang12}, superconductivity and electronic nematic order~\cite{ando2002electrical, daou2010broken, Achkar16, lawler2010intra, fujita2011spectroscopic} are intertwined, often co-existing and competing~\cite{Fradkin2015, Keimer2015, Kivelson03}. How these orders manifest and relate to each other depends on a number of factors including the doping, the level of disorder and the crystalline structure (tetragonal or orthorhombic, single layer or bilayer, lattice constants, etc). An approach to vary the crystalline structure, while keeping doping and disorder constant, is the application of uniaxial stress.  Examination of the subsequent response of charge density wave (CDW), nematic, and superconductivity orders to uniaxial stress provides a powerful pathway to understand the relationship of these intertwined orders.

Application of uniaxial stress in the cuprates has been shown to result in significant changes in the superconducting transition temperature and spin or charge density wave orders~\cite{ARUMUGAM2000,Takeshita04,Sasagawa05,Guguchia20,KIm18,Kim21,Choi22, Boyle2021,Wahlberg21,guguchia2023designing}. In the stripe-ordered cuprates La$_{1.74}$Eu$_{0.2}$Sr$_{0.16}$CuO$_4$ and La$_{1.48}$Nd$_{0.4}$Sr$_{0.12}$CuO$_4$, uniaxial stress applied in the $ab$ plane enhances the superconducting transition temperature, $T_C$~\cite{Takeshita04,ARUMUGAM2000}. More recent work in La$_{1.885}$Ba$_{0.115}$CuO$_4$ (LBCO), showed a similar enhancement of $T_C$, along with an accompanying suppression of magnetic order, as measured from muon spin rotation ($\mu $SR) spectroscopy~\cite{guguchia2023designing,Guguchia20}. This result is indicative of a competition between magnetic order and superconductivity that can be tuned by uniaxial stress.

Other studies have examined the impact of uniaxial stress on charge density wave order. In YBa$_2$Cu$_3$O$_{6+x}$ (YBCO), $\sim$1\% compressive uniaxial strain enhances (2D) CDW order and results in a unidirectional (3D) CDW order \cite{KIm18,Kim21}. An alternate approach in YBa$_2$Cu$_3$O$_{6+x}$ showed that tensile strain induced in thin films grown on an orthorhombic substrate results in a  suppression of CDW order perpendicular to the strain~\cite{Wahlberg21}. However, in La$_{1.875}$Sr$_{0.125}$CuO$_4$ (LSCO), uniaxial stress was found to enhance CDW order propagating perpendicular to the applied stress and suppress CDW order propagating parallel to the applied stress~\cite{Choi22,Wang22}.

How uniaxial stress impacts CDW order depends on details of a materials crystals structure. For instance, in LSCO, the structure is in the low-temperature orthorhombic (LTO) phase, characterized by tilts of CuO$_6$ octahedra about an axis diagonal to the Cu-O bond. For unidirectional CDW order propagating approximately along [100] and [010] (parallel to the Cu-O bond), the structure does not establish a preferred direction, providing an opportunity for the formation of domains of both [100]- and [010]-oriented CDW order~\cite{LSCO_Frison,LSCO_Vaknin}. Consequently, the application strain along [100] was interpreted to detwin the CDW order to favor only domains of unidirectional CDW order propagating along [100] for stress applied along [010]~\cite{Choi22}.

    \begin{figure}[ht]
     \begin{center}
   \resizebox{\columnwidth}{!}{\includegraphics{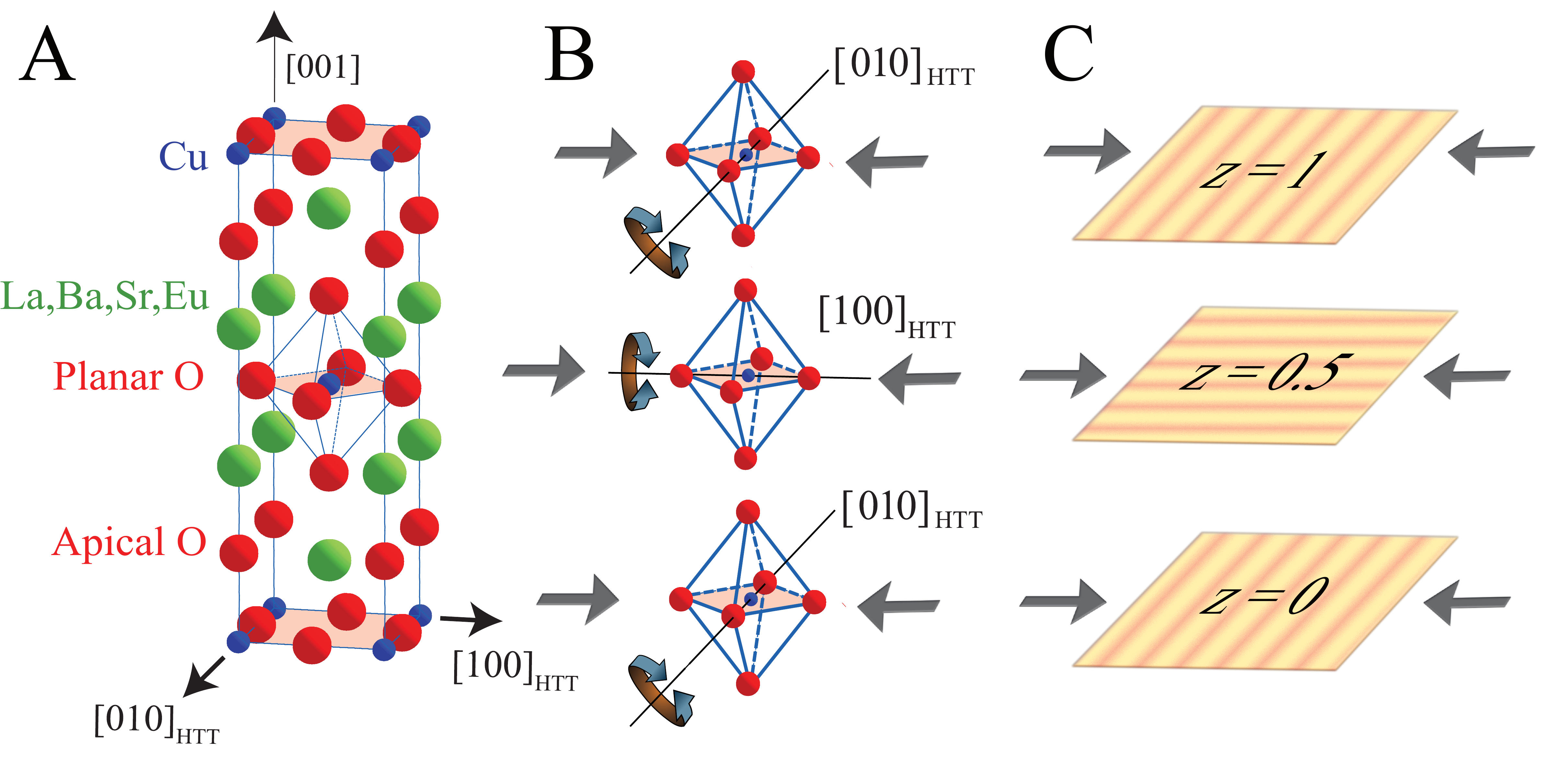}}
        \caption{\textbf{A} The crystal structure of (La,M)$_2$CuO$_4$ in the high temperature tetragonal (HTT) phase. Two CuO$_2$ planes at $z$~=~0 and $z$~=~0.5 are present within a single unit cell. \textbf{B} In the low temperature tetragonal (LTT) phase CuO$_6$ octahedra tilt along the Cu-O bond direction, with the tilt axis alternating between $a$ and $b$ for neighbouring layers. Uniaxial stress along $a$ will be parallel to the tilt axis for half the layers and perpendicular to the tilt axis for the other half of the layers. \textbf{C} The orientation of unidirectional stripe order alternates between neighbouring layers. Uniaxial stress along $a$ will be parallel or perpendicular to the CDW propagation wavevector, depending on the layer.
        \label{fig:fig1}}
        \end{center}
    \end{figure}

However, in many stripe-ordered cuprates, the application of strain may play a different role than in LSCO. In (La,M)$_2$CuO$_4$ substituted with larger rare earth ions Nd, Eu or Ba, a low temperature tetragonal (LTT) phase can be induced that is characterized by octahedral tilt axes parallel to the Cu-O bonds~\cite{Axe89,Suzuki89}. Numerous studies have recognized that the LTT structure stabilizes stripe order, with several compounds exhibiting stripe order to onset at the LTO to LTT phase transition~\cite{tranquada1995evidence,Hucker11,Hucker12}. The reason for this association is that the LTT structure, shown in Figure 1, induces anisotropy in the electronic structure, such as the hopping between Cu and O ($t_{pd}$), the nearest neighbour exchange interaction, $J$, or the charge transfer energy $\Delta_{pd}$~\cite{Yamase2000,Kampf01}. This anisotropy is thought to stabilize unidirectional stripe order that, within an individual layer, is oriented parallel or perpendicular to the octahedral tilt axis. Importantly, although each individual layer breaks $C_4$ rotational symmetry due to the octahedral tilts, the axis by which the octahedra tilt rotates by 90$^{\circ}$ between neighbouring CuO$_2$ planes such that the crystal structure remains tetragonal, as depicted in Figure 1B. Consequently, unidirectional spin and charge order in neighbouring layers alternates between propagating along [100] and [010]~\cite{tranquada1995evidence,Sears2023}, as depicted in Figure 1C.

Unlike in LSCO without Nd, Eu or Ba substitution, the structural distortion of the LTT phase serves to detwin the CDW order within an individual layer. As such, the application of uniaxial stress may be used to explore the role of anisotropy of the electronic structure on CDW order, rather than the balance of population of [100] and [010] domains within a layer. As shown in Figure 1, the application of a uniaxial stress along the [100] direction will act along the tilt axis for half of the layers and perpendicular to the tilt axis for the other half of the layers, affecting the anisotropy of the electronic structure differently for the two orientations of layers.

In this study we utilize resonant soft x-ray scattering to study the impact of uniaxial strain on both the structural phase transition and the CDW order in La$_{1.475}$Nd$_{0.4}$Sr$_{0.125}$CuO$_4$ (LNSCO). We find that the application of stress along [100] reduces the intensity of the CDW peak at \textbf{$Q$}$_{\textrm{CDW}}$ = ($-$0.24~0~1.5) by a factor of $\sim 2$, while having little impact on its correlation length or temperature dependence. In contrast, the applied [100] stress has only modest impact on the intensity of the CDW peak at (0~$-$0.24~1.5). The overall suppression of CDW order due to uniaxial stress is consistent with competition between CDW order and superconductivity. More specifically, uniaxial stress along [100] modifies the anisotropy of the electronic structure in a manner that suppresses CDW order, and thus in turn enables the enhancement of the superconducting transition temperature \cite{Takeshita04,Guguchia20,guguchia2023designing}.

Resonant x-ray scattering measurements presented in this study were performed on a cut, polished cuboidal sample of La$_{1.475}$Nd$_{0.4}$Sr$_{0.125}$CuO$_4$. The CDW ($-$0.24 0 1.5) and (0 $-$0.24 1.5) Bragg peaks were investigated at a photon energy corresponding to the peak of the Cu-$L_3$ absorption edge (931.3 eV). Note, measurements of the CDW peaks are shown as raw data, normalized only to the incident beam intensity, but otherwise free of any background subtraction. Measurements were found to be reproducible to an accuracy of $\sim2\%$ of the total signal. In addition, to examine the LTO to LTT structural phase transition, we measured the (0~0~1) Bragg peak beak at the Cu-$L$ edge (931.3 eV) and at an energy associated with the apical oxygen (533.3 eV). 

Using a custom uniaxial stress device described in the supplementary information~\cite{suppinfo}, uniaxial stress was applied parallel to [100] axis of the high-temperature tetragonal (HTT) unit cell.  The sample was measured both unstrained and with applied uniaxial stress. The magnitude of the applied stress imparted to the sample is not characterized in the device. However, the maximum strain is estimated to be of less than 0.2$\%$ and the stress less than $< 0.5$ GPa (see supplementary information~\cite{suppinfo}). With applied stress, the CDW Bragg peaks were measured for CDW order propagating both parallel and perpendicular to the applied stress.

    \begin{figure}[th]
     \resizebox{\columnwidth}{!}{\includegraphics{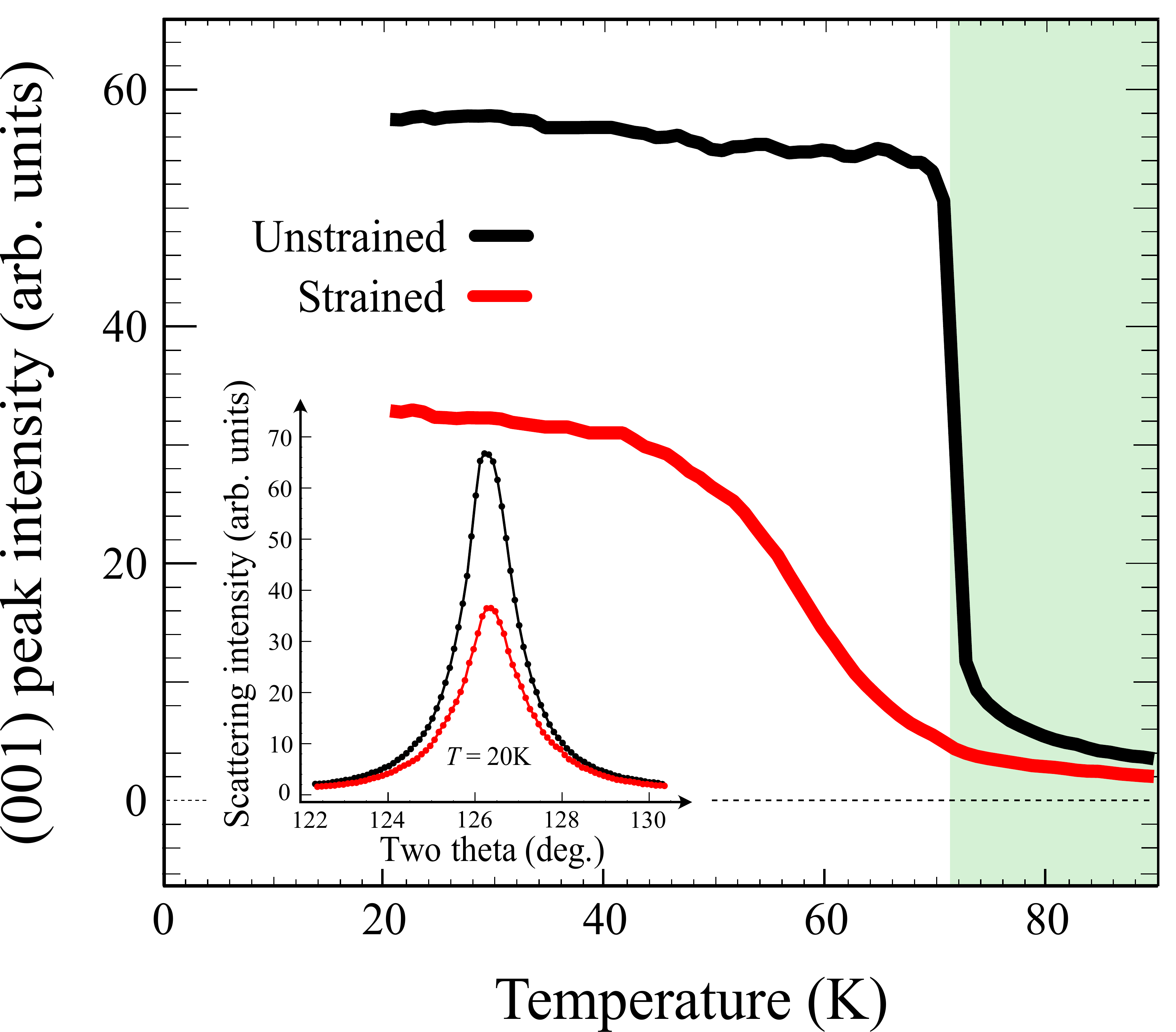}}
        \caption{Temperature dependence of the (0~0~1) Bragg peak at apical O-$K$ edge under varying strain (increasing as shown by an arrow). Scattering intensity is normalized at 20 K. The inset shows a theta-2theta scan of the (0~0~1) Bragg peak on-resonance at the apical O-$K$ edge, comparing the strained (red) and the unstrained (black) case. Notably, the peak intensity drops upon applying uniaxial stress. The green-shaded region (above 70 K) depicts the LTO phase in the unstrained case.
        \label{fig:fig2}}
    \end{figure}
    
First we examine the impact of uniaxial stress on the structure. As shown in Figure~2, the temperature dependence of the (0~0~1) peak measured at the apical O-$K$ edge in an unstrained configuration, exhibits a sharp transition to the LTT phase at 70 K, consistent with previous unstrained measurements on this sample \cite{Achkar16,Gupta21}. The application of stress along [100] results in a decrease in the (0~0~1) peak intensity at low temperatures, as well as, a more gradual temperature dependence, indicating a change in the structure under uniaxial stress. Details of the structural changes induced by uniaxial stress cannot be fully resolved by measuring a single Bragg peak. However, since the (0~0~1) peak results from a difference in the orbital symmetry of apical O atoms between neighbouring (La,M)$_2$O$_2$ layers~\cite{Fink11,Wilkins11,Achkar16}, the more gradual temperature dependence of the (0~0~1) peak under applied stress is understood to be associated with a reduction in the difference in orbital symmetry between neighbouring layers. This reduction may result from changes in the angles by which CuO$_6$ octahedra tilt out of the $ab$ plane and/or changes in the octahedral tilt axis away from the Cu-O bond direction.
 
Notably, the strain dependence of the (0~0~1) Bragg peak in Figure 2 is in qualitative agreement with recent reports hard x-ray scattering study of the LTO-LTT transition under unaxial stress in another stripe-ordered cuprate La$_{1.885}$Ba$_{0.115}$CuO$_4$~\cite{guguchia2023designing}, albeit with stress applied along [110] instead of along [100]. In that study, Guguchia~$et~al$.~\cite{guguchia2023designing} show the LTT transition to be completely suppressed for compressive uniaxial stress above {$\sigma_{[110]}\sim$0.06 GPa}. However, for lower stress values ($\sigma_{[110]}\sim$0.017 GPa), the LTT phase remains, but with the LTT Bragg peak intensity suppressed, the onset temperature only weakly dependent on stress, and the LTO-LTT transition broadened~\cite{guguchia2023designing}, qualitatively similar to the dependence on uniaxial stress dependence of the (0~0~1) we observe in LNSCO for stress applied along [100] in the range of pressure applied.
    \begin{figure}[ht]
     \resizebox{\columnwidth}{!}{\includegraphics{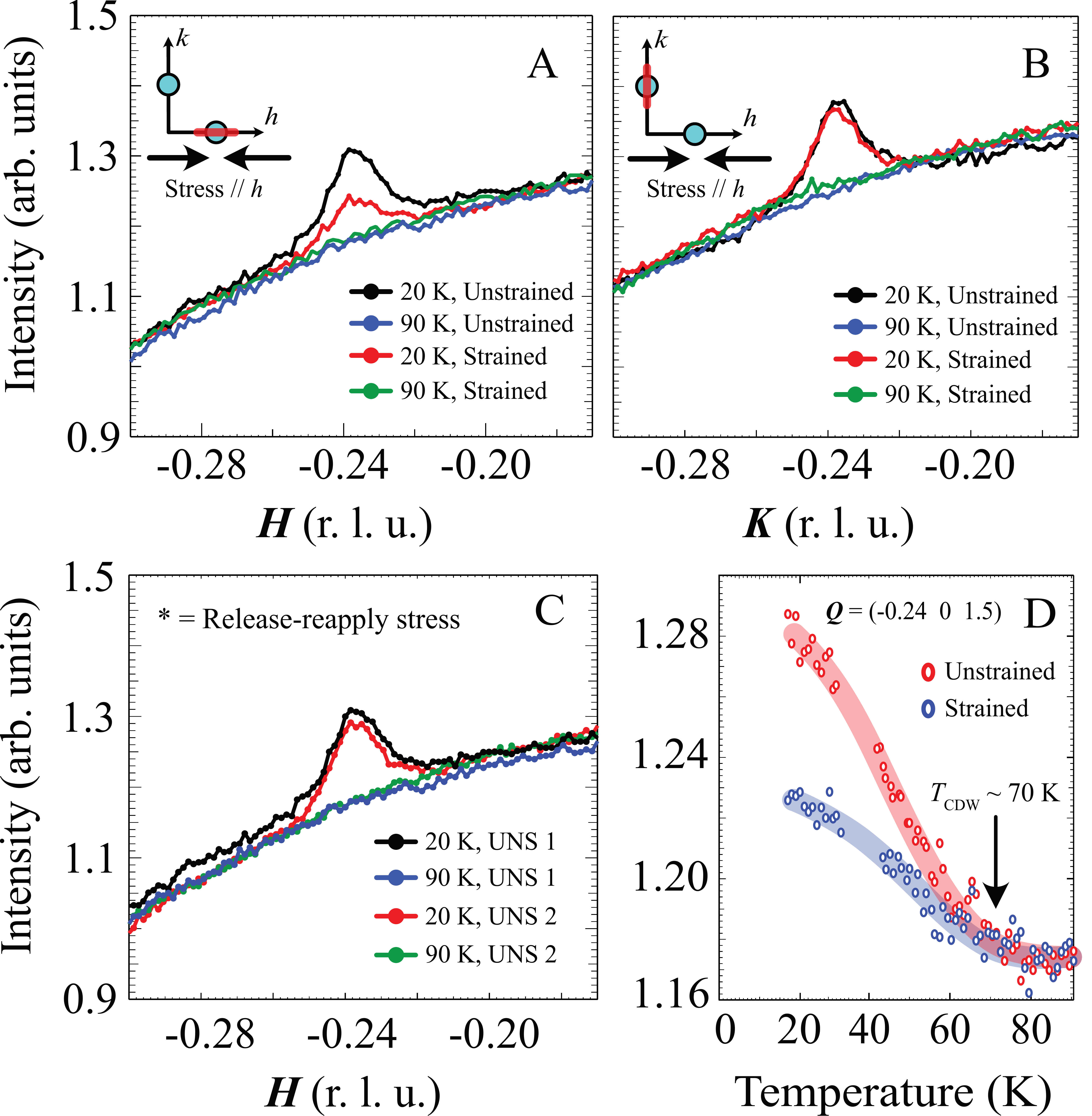}}
        \caption{Response of CDW Bragg peaks to applied uniaxial stress along [100]. \textbf{A}) and \textbf{B}) Intensity at 20 K and at 90 K, above the CDW ordering temperature, for the unstrained and strained case. In \textbf{A})\ measurements are shown through the Bragg peak ($-$0.24~0~1.5) peak position, corresponding to CDW order propagating parallel to the applied stress.  In \textbf{B} measurements are shown  through the Bragg peak (0~$-$0.24~1.5) peak position, corresponding to CDW order propagating perpendicular to the applied stress. \textbf{C}) Measurements through ($-$0.24~0~1.5) in the unstrained configuration, before (UNS 1) and after releasing (UNS 2) applied stress to the sample.  \textbf{D}) The temperature dependence at ($-$0.24~0~1.5) with and before strain (UNS1). The onset temperature of the CDW order is similar in the strained and unstrained configurations. All panels show raw data. 
        \label{fig:fig3}}
    \end{figure}

In figure 3, we show the impact of uniaxial stress on CDW order. When the sample is unstrained, we observe comparable peak intensities for the \mbox{($-$0.24 0 1.5)} and \mbox{(0 $-$0.24 1.5)} peaks. The response of CDW order to uniaxial stress, however, is asymmetric. Whereas uniaxial stress along [100] reduces the intensity of the \mbox{($-$0.24~0~1.5)} Bragg peak by a factor of $\sim 2$ at 20~K, this same stress has minimal impact on the intensity of the \mbox{(0~$-$0.24~1.5)} Bragg peak. Moreover, although the intensity of the \mbox{($-$0.24~0~1.5)} peak decreases with the application of stress, the width of the Bragg peak, associated with the CDW correlation length along the $a$-axis, is unaffected. 

In both cases, the application of uniaxial stress does not change the background measured at 90~K, which is dominated by x-ray fluorescence from the sample as detailed in the supplementary information~\cite{suppinfo}. Charge density fluctuations at high temperatures, as have been reported in other studies~\cite{Arpaia21,Boyle2021}, may also occur but cannot be resolved unambiguously in this set of measurements.
    \begin{figure}[ht]
    \begin{center}
   \resizebox{\columnwidth}{!}{\includegraphics{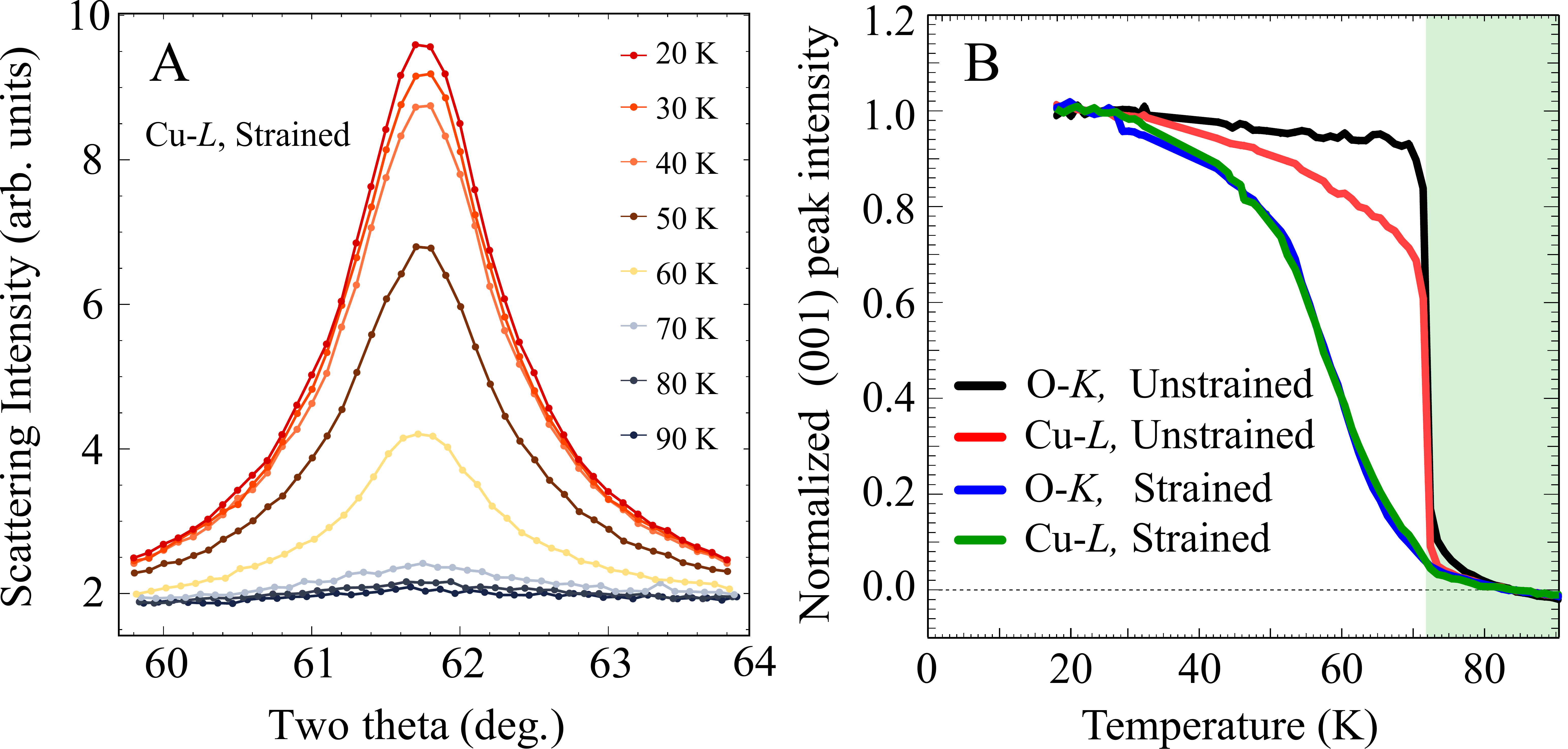}}
        \caption{Temperature dependence of the structural-nematic (0~0~1) Bragg peaks for planar and apical atoms. \textbf{A} Temperature evolution of the theta-2theta scans of the (0~0~1) Bragg peak at the Cu-$L$ edge under in-plane, compressive uniaxial strain. \textbf{B} Comparison of temperature dependence of the (0~0~1) peak measured at the Cu-$L$ (931.3 eV) and apical O-$K$ edge (533.3 eV) energies for unstrained and strained case. Scattering intensity is normalized at 20 K. The temperature evolution of the (0~0~1) peak at the two energies differs in the unstrained case~\cite{Achkar16}, but evolves similarly under uniaxial stress.
        \label{fig:fig4}}
        \end{center}
    \end{figure}

The temperature dependence of the intensity at \mbox{($-$0.24~0~1.5)} in strained and unstrained configurations is shown in Figure~3D. This shows that the onset temperature is not strongly impacted by strain (we are unable to clearly resolve a change in $T_{\textrm{CDW}}$), with the main impact of strain being the reduction in peak intensity in the region where CDW order exists.

To verify that the strain-induced suppression in the ($-$0.24~0~1.5) peak intensity is not associated with an irreversible degradation of the sample (cracking, buckling, etc.) upon straining, we released the stress and remeasured the sample in an unstrained configuration (UNS~2). These measurements show a recovery of the original CDW peak intensity (see UNS~1 in Figure~3C), as well as a recovery of the (0~0~1) peak temperature dependence.

Notably, our results with uniaxial compressive stress differ from reported measurements in LNSCO with uniaxial tensile stress by Boyle~$et~al.$~\cite{Boyle2021}. They also observe a reduction in the CDW peak intensity at low $T$ under stress. However, where we find $T_{\textrm{LTT}}$ and $T_{\textrm{CDW}}$ are unchanged by the compressive stress we applied, they report that a tensile strain of $\varepsilon_a = +0.046 \pm 0.026\%$ reduced both $T_{\textrm{CDW}}$ and $T_{\textrm{LTT}}$ by 29 K. Moreover, they do not observe an appreciable broadening of the LTT transition or reduction in the (0~0~1) peak intensity at low temperatures, as we observed in our study. This may indicate a marked difference in the impact of tensile versus compressive stress or in the magnitudes of stress applied in the two studies, warranting further investigation.

The compressive stress dependence observed in our study is also in contrast to that of reports in LSCO, where CDW order occurs within the LTO structural phase. In LSCO, CDW propagating along the $a$ and $b$ axes would be degenerate in an unstrained crystal, giving rise to domains of unidirectional CDW order that run along both $a$ and $b$ within an individual layer~\cite{Choi22}. As such, even a small uniaxial strain along $a$ or $b$ can break the degeneracy, detwinning the CDW such that only $a$ or $b$ oriented CDW order occur, consistent with the measurements of Choi $et~al.$~\cite{Choi22}. Notably, in LSCO application of strain beyond that required to detwin the CDW order did not further enhance the CDW order~\cite{Choi22}. Connecting the observations to our results in LNSCO, the application of uniaxial strain here does not have a similar effect of detwinning CDW order -- enhancing CDW order along $b$ and suppressing CDW order along $a$. This is likely due to octahedral tilts along $a$ and $b$ in the LTT structure already being effective at detwinning the CDW order within individual layers.

In our study, uniaxial stress modifies the octahedral distortions characteristic of the LTT phase, which subsequently impacts the in-plane anisotropy of the electronic structure within an individual layer, such as the nearest-neighbour hopping and exchange interactions. This change in the electronic anisotropy, in turn, can affect the amplitude of CDW order. Structural refinements of CDW order in the LTT phase of LBCO have identified that CDW order within an individual layer propagates along a direction parallel to the bent O-Cu-O bonds as depicted in Figure 1~\cite{Sears2023}. While it is not clear how the electronic anisotropy of the individual layers changes in response to anisotropic strain, one might expect that the electronic anisotropy increases for the layers with the applied stress parallel to the straight O-Cu-O bonds and decreases for layers with the applied stress perpendicular to the straight O-Cu-O bonds. If this is indeed the case, our findings of a suppression of CDW order along $H$ (parallel to the applied stress) indicate that a reduction in electronic anisotropy of a layer results in a reduction of CDW order, but the converse need not be the case.

Ultimately, the net effect of applying uniaxial stress along the Cu-O bond direction is to suppress CDW order. This is in accordance with observed increases in the superconducting transition temperature by applying stress in the [100] direction in LNSCO \cite{Takeshita04} and related systems La$_{1.65}$Eu$_{0.2}$Sr$_{0.15}$CuO$_4$~\cite{Takeshita04} and La$_{1.885}$Ba$_{0.115}$CuO$_4$~\cite{Guguchia20} that share the LTT structure with LNSCO. As such, at least part of the increase in $T_c$ may be attributed to the competition of CDW order with superconductivity. Namely, uniaxial stress affects the anisotropy of the electronic structure in a manner that suppresses CDW order, resulting in an enhancement of superconductivity. 

In addition to investigating CDW order, we also probe the $Q_x = Q_y = 0$ electronic anisotropy by measuring the temperature dependence of the (0~0~1) Bragg peak under uniaxial strain at the Cu-$L$ edge, in addition to O-$K$ edge (at an energy corresponding to apical oxygen).\cite{Achkar16,Gupta21,bluschke2022} As shown in Figure~4, the temperature dependence of the (0~0~1) peak intensity is different and more gradual when measured at the Cu-$L$ relative to the O-$K$ edge, consistent with previous studies without strain~\cite{Achkar16,Gupta21}. However, uniaxial stress appears to eliminate this difference in the $T$ dependence.

Achkar $et~al.$ \cite{Achkar16} argued that this difference in temperature dependence is due to electronic nematic order that is coupled to CDW order and is in addition electronic asymmetry that directly results from the structure distortions. Moreover, although the (0~0~1) peak measures the difference in orbital symmetry between neighbouring layers, in the unstrained case the symmetry of the crystal structure is such that this difference in the interlayer orbital asymmetry maps to a measure of nematicity within a single layer. However, because of the inequivalence of neighbouring layers under anisotropic strain, this mapping is not valid, complicating the interpretation of the (0~0~1) peak $T$-dependence under uniaxial stress.  

The lack of a difference in the $T$-dependence between Cu-$L$ and apical O-$K$ measurements may indicate that uniaxial stress suppresses electronic nematic order within individual layers. However, it may also indicate that nematic order remains strong and is perhaps saturated.  In this scenario, uniaxial stress may align electronic nematic order between neighbouring layers, such that signatures of it cancel in measurements of the (0~0~1) Bragg peak. Future work will be required to differentiate the origin of this anomalous result.

In conclusion, we find that in La$_{1.475}$Nd$_{0.4}$Sr$_{0.125}$CuO$_4$, uniaxial stress along [100] suppresses the magnitude of CDW order parallel to the applied stress, but has little impact on CDW order propagating perpendicular to the applied stress.  We attribute this suppression to modification of the electronic asymmetry within the CuO$_2$ planes rather than a detwinning of CDW order, seen in LSCO.  Signatures of nematic order, as observed via the relative temperature dependence of the (0~0~1) peak at the Cu-$L$ edge and O-$K$ edge, are also observed to be suppressed by uniaxial strain.  This suppression of CDW order is likely linked to the enhancement of superconductivity under uniaxial stress along [100] direction.

\begin{acknowledgments}
We gratefully acknowledge Christopher Keegan, Erica Carlson, Steve Kivelson, and Louis Taillefer for useful discussions. We also acknowledge assistance from Rafael Mirabal and Michael Boulaine during the sample etching process. This research was supported by the Natural Sciences and Engineering Research Council (NSERC). N.K. Gupta acknowledges support from the Waterloo Institute of Nanotechnology (WIN).  Part of the research described in this paper was performed at the Canadian Light Source, a national research facility of the University of Saskatchewan, which is supported by the Canada Foundation for Innovation (CFI), the Natural Sciences and Engineering Research Council (NSERC), the National Research Council Canada (NRC), the Canadian Institutes of Health Research (CIHR), the Government of Saskatchewan, and the University of Saskatchewan.
\end{acknowledgments}

%

\begin{thebibliography}{37}%
\makeatletter
\providecommand \@ifxundefined [1]{%
 \@ifx{#1\undefined}
}%
\providecommand \@ifnum [1]{%
 \ifnum #1\expandafter \@firstoftwo
 \else \expandafter \@secondoftwo
 \fi
}%
\providecommand \@ifx [1]{%
 \ifx #1\expandafter \@firstoftwo
 \else \expandafter \@secondoftwo
 \fi
}%
\providecommand \natexlab [1]{#1}%
\providecommand \enquote  [1]{``#1''}%
\providecommand \bibnamefont  [1]{#1}%
\providecommand \bibfnamefont [1]{#1}%
\providecommand \citenamefont [1]{#1}%
\providecommand \href@noop [0]{\@secondoftwo}%
\providecommand \href [0]{\begingroup \@sanitize@url \@href}%
\providecommand \@href[1]{\@@startlink{#1}\@@href}%
\providecommand \@@href[1]{\endgroup#1\@@endlink}%
\providecommand \@sanitize@url [0]{\catcode `\\12\catcode `\$12\catcode
  `\&12\catcode `\#12\catcode `\^12\catcode `\_12\catcode `\%12\relax}%
\providecommand \@@startlink[1]{}%
\providecommand \@@endlink[0]{}%
\providecommand \url  [0]{\begingroup\@sanitize@url \@url }%
\providecommand \@url [1]{\endgroup\@href {#1}{\urlprefix }}%
\providecommand \urlprefix  [0]{URL }%
\providecommand \Eprint [0]{\href }%
\providecommand \doibase [0]{https://doi.org/}%
\providecommand \selectlanguage [0]{\@gobble}%
\providecommand \bibinfo  [0]{\@secondoftwo}%
\providecommand \bibfield  [0]{\@secondoftwo}%
\providecommand \translation [1]{[#1]}%
\providecommand \BibitemOpen [0]{}%
\providecommand \bibitemStop [0]{}%
\providecommand \bibitemNoStop [0]{.\EOS\space}%
\providecommand \EOS [0]{\spacefactor3000\relax}%
\providecommand \BibitemShut  [1]{\csname bibitem#1\endcsname}%
\let\auto@bib@innerbib\@empty
\bibitem [{\citenamefont {Tranquada}\ \emph {et~al.}(1995)\citenamefont
  {Tranquada}, \citenamefont {Sternlieb}, \citenamefont {Axe}, \citenamefont
  {Nakamura},\ and\ \citenamefont {Uchida}}]{tranquada1995evidence}%
  \BibitemOpen
  \bibfield  {author} {\bibinfo {author} {\bibfnamefont {J.}~\bibnamefont
  {Tranquada}}, \bibinfo {author} {\bibfnamefont {B.}~\bibnamefont
  {Sternlieb}}, \bibinfo {author} {\bibfnamefont {J.}~\bibnamefont {Axe}},
  \bibinfo {author} {\bibfnamefont {Y.}~\bibnamefont {Nakamura}},\ and\
  \bibinfo {author} {\bibfnamefont {S.}~\bibnamefont {Uchida}},\ }\bibfield
  {title} {\bibinfo {title} {Evidence for stripe correlations of spins and
  holes in copper oxide superconductors},\ }\href
  {https://doi.org/https://doi.org/10.1038/375561a0} {\bibfield  {journal}
  {\bibinfo  {journal} {Nature}\ }\textbf {\bibinfo {volume} {375}},\ \bibinfo
  {pages} {561} (\bibinfo {year} {1995})}\BibitemShut {NoStop}%
\bibitem [{\citenamefont {Ghiringhelli}\ \emph {et~al.}(2012)\citenamefont
  {Ghiringhelli}, \citenamefont {Le~Tacon}, \citenamefont {Minola},
  \citenamefont {Blanco-Canosa}, \citenamefont {Mazzoli}, \citenamefont
  {Brookes}, \citenamefont {De~Luca}, \citenamefont {Frano}, \citenamefont
  {Hawthorn}, \citenamefont {He}, \citenamefont {Loew}, \citenamefont {Sala},
  \citenamefont {Peets}, \citenamefont {Salluzzo}, \citenamefont {Schierle},
  \citenamefont {Sutarto}, \citenamefont {Sawatzky}, \citenamefont {Weschke},
  \citenamefont {Keimer},\ and\ \citenamefont {Braicovich}}]{Ghiringhelli12}%
  \BibitemOpen
  \bibfield  {author} {\bibinfo {author} {\bibfnamefont {G.}~\bibnamefont
  {Ghiringhelli}}, \bibinfo {author} {\bibfnamefont {M.}~\bibnamefont
  {Le~Tacon}}, \bibinfo {author} {\bibfnamefont {M.}~\bibnamefont {Minola}},
  \bibinfo {author} {\bibfnamefont {S.}~\bibnamefont {Blanco-Canosa}}, \bibinfo
  {author} {\bibfnamefont {C.}~\bibnamefont {Mazzoli}}, \bibinfo {author}
  {\bibfnamefont {N.~B.}\ \bibnamefont {Brookes}}, \bibinfo {author}
  {\bibfnamefont {G.~M.}\ \bibnamefont {De~Luca}}, \bibinfo {author}
  {\bibfnamefont {A.}~\bibnamefont {Frano}}, \bibinfo {author} {\bibfnamefont
  {D.~G.}\ \bibnamefont {Hawthorn}}, \bibinfo {author} {\bibfnamefont
  {F.}~\bibnamefont {He}}, \bibinfo {author} {\bibfnamefont {T.}~\bibnamefont
  {Loew}}, \bibinfo {author} {\bibfnamefont {M.~M.}\ \bibnamefont {Sala}},
  \bibinfo {author} {\bibfnamefont {D.~C.}\ \bibnamefont {Peets}}, \bibinfo
  {author} {\bibfnamefont {M.}~\bibnamefont {Salluzzo}}, \bibinfo {author}
  {\bibfnamefont {E.}~\bibnamefont {Schierle}}, \bibinfo {author}
  {\bibfnamefont {R.}~\bibnamefont {Sutarto}}, \bibinfo {author} {\bibfnamefont
  {G.~A.}\ \bibnamefont {Sawatzky}}, \bibinfo {author} {\bibfnamefont
  {E.}~\bibnamefont {Weschke}}, \bibinfo {author} {\bibfnamefont
  {B.}~\bibnamefont {Keimer}},\ and\ \bibinfo {author} {\bibfnamefont
  {L.}~\bibnamefont {Braicovich}},\ }\bibfield  {title} {\bibinfo {title}
  {Long-range incommensurate charge fluctuations in
  {(Y,Nd)Ba$_2$Cu$_3$O$_{6+x}$}},\ }\href
  {https://doi.org/10.1126/science.1223532} {\bibfield  {journal} {\bibinfo
  {journal} {Science}\ }\textbf {\bibinfo {volume} {337}},\ \bibinfo {pages}
  {821} (\bibinfo {year} {2012})}\BibitemShut {NoStop}%
\bibitem [{\citenamefont {Chang}\ \emph {et~al.}(2012)\citenamefont {Chang},
  \citenamefont {Blackburn}, \citenamefont {Holmes}, \citenamefont
  {Christensen}, \citenamefont {Larsen}, \citenamefont {Mesot}, \citenamefont
  {Liang}, \citenamefont {Bonn}, \citenamefont {Hardy}, \citenamefont
  {Watenphul}, \citenamefont {von Zimmermann}, \citenamefont {Forgan},\ and\
  \citenamefont {Hayden}}]{Chang12}%
  \BibitemOpen
  \bibfield  {author} {\bibinfo {author} {\bibfnamefont {J.}~\bibnamefont
  {Chang}}, \bibinfo {author} {\bibfnamefont {E.}~\bibnamefont {Blackburn}},
  \bibinfo {author} {\bibfnamefont {A.~T.}\ \bibnamefont {Holmes}}, \bibinfo
  {author} {\bibfnamefont {N.~B.}\ \bibnamefont {Christensen}}, \bibinfo
  {author} {\bibfnamefont {J.}~\bibnamefont {Larsen}}, \bibinfo {author}
  {\bibfnamefont {J.}~\bibnamefont {Mesot}}, \bibinfo {author} {\bibfnamefont
  {R.}~\bibnamefont {Liang}}, \bibinfo {author} {\bibfnamefont {D.~A.}\
  \bibnamefont {Bonn}}, \bibinfo {author} {\bibfnamefont {W.~N.}\ \bibnamefont
  {Hardy}}, \bibinfo {author} {\bibfnamefont {A.}~\bibnamefont {Watenphul}},
  \bibinfo {author} {\bibfnamefont {M.}~\bibnamefont {von Zimmermann}},
  \bibinfo {author} {\bibfnamefont {E.~M.}\ \bibnamefont {Forgan}},\ and\
  \bibinfo {author} {\bibfnamefont {S.~M.}\ \bibnamefont {Hayden}},\ }\bibfield
   {title} {\bibinfo {title} {{Direct observation of competition between
  superconductivity and charge density wave order in
  YBa$_2$Cu$_3$O$_{6.67}$}},\ }\href {https://doi.org/10.1038/nphys2456}
  {\bibfield  {journal} {\bibinfo  {journal} {Nat. Phys.}\ }\textbf {\bibinfo
  {volume} {8}},\ \bibinfo {pages} {871} (\bibinfo {year} {2012})}\BibitemShut
  {NoStop}%
\bibitem [{\citenamefont {Ando}\ \emph {et~al.}(2002)\citenamefont {Ando},
  \citenamefont {Segawa}, \citenamefont {Komiya},\ and\ \citenamefont
  {Lavrov}}]{ando2002electrical}%
  \BibitemOpen
  \bibfield  {author} {\bibinfo {author} {\bibfnamefont {Y.}~\bibnamefont
  {Ando}}, \bibinfo {author} {\bibfnamefont {K.}~\bibnamefont {Segawa}},
  \bibinfo {author} {\bibfnamefont {S.}~\bibnamefont {Komiya}},\ and\ \bibinfo
  {author} {\bibfnamefont {A.~N.}\ \bibnamefont {Lavrov}},\ }\bibfield  {title}
  {\bibinfo {title} {Electrical resistivity anisotropy from self-organized one
  dimensionality in high-temperature superconductors},\ }\href
  {https://doi.org/10.1103/PhysRevLett.88.137005} {\bibfield  {journal}
  {\bibinfo  {journal} {Phys. Rev. Lett.}\ }\textbf {\bibinfo {volume} {88}},\
  \bibinfo {pages} {137005} (\bibinfo {year} {2002})}\BibitemShut {NoStop}%
\bibitem [{\citenamefont {Daou}\ \emph {et~al.}(2010)\citenamefont {Daou},
  \citenamefont {Chang}, \citenamefont {LeBoeuf}, \citenamefont
  {Cyr-Choiniere}, \citenamefont {Lalibert{\'e}}, \citenamefont
  {Doiron-Leyraud}, \citenamefont {Ramshaw}, \citenamefont {Liang},
  \citenamefont {Bonn}, \citenamefont {Hardy} \emph {et~al.}}]{daou2010broken}%
  \BibitemOpen
  \bibfield  {author} {\bibinfo {author} {\bibfnamefont {R.}~\bibnamefont
  {Daou}}, \bibinfo {author} {\bibfnamefont {J.}~\bibnamefont {Chang}},
  \bibinfo {author} {\bibfnamefont {D.}~\bibnamefont {LeBoeuf}}, \bibinfo
  {author} {\bibfnamefont {O.}~\bibnamefont {Cyr-Choiniere}}, \bibinfo {author}
  {\bibfnamefont {F.}~\bibnamefont {Lalibert{\'e}}}, \bibinfo {author}
  {\bibfnamefont {N.}~\bibnamefont {Doiron-Leyraud}}, \bibinfo {author}
  {\bibfnamefont {B.}~\bibnamefont {Ramshaw}}, \bibinfo {author} {\bibfnamefont
  {R.}~\bibnamefont {Liang}}, \bibinfo {author} {\bibfnamefont
  {D.}~\bibnamefont {Bonn}}, \bibinfo {author} {\bibfnamefont {W.}~\bibnamefont
  {Hardy}}, \emph {et~al.},\ }\bibfield  {title} {\bibinfo {title} {Broken
  rotational symmetry in the pseudogap phase of a high-${T}_{c}$
  superconductor},\ }\href@noop {} {\bibfield  {journal} {\bibinfo  {journal}
  {Nature}\ }\textbf {\bibinfo {volume} {463}},\ \bibinfo {pages} {519}
  (\bibinfo {year} {2010})}\BibitemShut {NoStop}%
\bibitem [{\citenamefont {Achkar}\ \emph
  {et~al.}(2016{\natexlab{a}})\citenamefont {Achkar}, \citenamefont {Zwiebler},
  \citenamefont {McMahon}, \citenamefont {He}, \citenamefont {Sutarto},
  \citenamefont {Djianto}, \citenamefont {Hao}, \citenamefont {Gingras},
  \citenamefont {Hücker}, \citenamefont {Gu}, \citenamefont {Revcolevschi},
  \citenamefont {Zhang}, \citenamefont {Kim}, \citenamefont {Geck},\ and\
  \citenamefont {Hawthorn}}]{Achkar16}%
  \BibitemOpen
  \bibfield  {author} {\bibinfo {author} {\bibfnamefont {A.~J.}\ \bibnamefont
  {Achkar}}, \bibinfo {author} {\bibfnamefont {M.}~\bibnamefont {Zwiebler}},
  \bibinfo {author} {\bibfnamefont {C.}~\bibnamefont {McMahon}}, \bibinfo
  {author} {\bibfnamefont {F.}~\bibnamefont {He}}, \bibinfo {author}
  {\bibfnamefont {R.}~\bibnamefont {Sutarto}}, \bibinfo {author} {\bibfnamefont
  {I.}~\bibnamefont {Djianto}}, \bibinfo {author} {\bibfnamefont
  {Z.}~\bibnamefont {Hao}}, \bibinfo {author} {\bibfnamefont {M.~J.~P.}\
  \bibnamefont {Gingras}}, \bibinfo {author} {\bibfnamefont {M.}~\bibnamefont
  {Hücker}}, \bibinfo {author} {\bibfnamefont {G.~D.}\ \bibnamefont {Gu}},
  \bibinfo {author} {\bibfnamefont {A.}~\bibnamefont {Revcolevschi}}, \bibinfo
  {author} {\bibfnamefont {H.}~\bibnamefont {Zhang}}, \bibinfo {author}
  {\bibfnamefont {Y.-J.}\ \bibnamefont {Kim}}, \bibinfo {author} {\bibfnamefont
  {J.}~\bibnamefont {Geck}},\ and\ \bibinfo {author} {\bibfnamefont {D.~G.}\
  \bibnamefont {Hawthorn}},\ }\bibfield  {title} {\bibinfo {title} {Nematicity
  in stripe-ordered cuprates probed via resonant x-ray scattering},\ }\href
  {https://doi.org/10.1126/science.aad1824} {\bibfield  {journal} {\bibinfo
  {journal} {Science}\ }\textbf {\bibinfo {volume} {351}},\ \bibinfo {pages}
  {576} (\bibinfo {year} {2016}{\natexlab{a}})}\BibitemShut {NoStop}%
\bibitem [{\citenamefont {Lawler}\ \emph {et~al.}(2010)\citenamefont {Lawler},
  \citenamefont {Fujita}, \citenamefont {Lee}, \citenamefont {Schmidt},
  \citenamefont {Kohsaka}, \citenamefont {Kim}, \citenamefont {Eisaki},
  \citenamefont {Uchida}, \citenamefont {Davis}, \citenamefont {Sethna} \emph
  {et~al.}}]{lawler2010intra}%
  \BibitemOpen
  \bibfield  {author} {\bibinfo {author} {\bibfnamefont {M.}~\bibnamefont
  {Lawler}}, \bibinfo {author} {\bibfnamefont {K.}~\bibnamefont {Fujita}},
  \bibinfo {author} {\bibfnamefont {J.}~\bibnamefont {Lee}}, \bibinfo {author}
  {\bibfnamefont {A.}~\bibnamefont {Schmidt}}, \bibinfo {author} {\bibfnamefont
  {Y.}~\bibnamefont {Kohsaka}}, \bibinfo {author} {\bibfnamefont {C.~K.}\
  \bibnamefont {Kim}}, \bibinfo {author} {\bibfnamefont {H.}~\bibnamefont
  {Eisaki}}, \bibinfo {author} {\bibfnamefont {S.}~\bibnamefont {Uchida}},
  \bibinfo {author} {\bibfnamefont {J.}~\bibnamefont {Davis}}, \bibinfo
  {author} {\bibfnamefont {J.}~\bibnamefont {Sethna}}, \emph {et~al.},\
  }\bibfield  {title} {\bibinfo {title} {Intra-unit-cell electronic nematicity
  of the high-{$T_{c}$} copper-oxide pseudogap states},\ }\href@noop {}
  {\bibfield  {journal} {\bibinfo  {journal} {Nature}\ }\textbf {\bibinfo
  {volume} {466}},\ \bibinfo {pages} {347} (\bibinfo {year}
  {2010})}\BibitemShut {NoStop}%
\bibitem [{\citenamefont {Fujita}\ \emph {et~al.}(2011)\citenamefont {Fujita},
  \citenamefont {R.~Schmidt}, \citenamefont {Kim}, \citenamefont {J.~Lawler},
  \citenamefont {Hai~Lee}, \citenamefont {Davis}, \citenamefont {Eisaki},\ and\
  \citenamefont {Uchida}}]{fujita2011spectroscopic}%
  \BibitemOpen
  \bibfield  {author} {\bibinfo {author} {\bibfnamefont {K.}~\bibnamefont
  {Fujita}}, \bibinfo {author} {\bibfnamefont {A.}~\bibnamefont {R.~Schmidt}},
  \bibinfo {author} {\bibfnamefont {E.-A.}\ \bibnamefont {Kim}}, \bibinfo
  {author} {\bibfnamefont {M.}~\bibnamefont {J.~Lawler}}, \bibinfo {author}
  {\bibfnamefont {D.}~\bibnamefont {Hai~Lee}}, \bibinfo {author} {\bibfnamefont
  {J.}~\bibnamefont {Davis}}, \bibinfo {author} {\bibfnamefont
  {H.}~\bibnamefont {Eisaki}},\ and\ \bibinfo {author} {\bibfnamefont {S.-i.}\
  \bibnamefont {Uchida}},\ }\bibfield  {title} {\bibinfo {title} {Spectroscopic
  imaging scanning tunneling microscopy studies of electronic structure in the
  superconducting and pseudogap phases of cuprate high-${T}_{c}$
  superconductors},\ }\href@noop {} {\bibfield  {journal} {\bibinfo  {journal}
  {Journal of the Physical Society of Japan}\ }\textbf {\bibinfo {volume}
  {81}},\ \bibinfo {pages} {011005} (\bibinfo {year} {2011})}\BibitemShut
  {NoStop}%
\bibitem [{\citenamefont {Fradkin}\ \emph {et~al.}(2015)\citenamefont
  {Fradkin}, \citenamefont {Kivelson},\ and\ \citenamefont
  {Tranquada}}]{Fradkin2015}%
  \BibitemOpen
  \bibfield  {author} {\bibinfo {author} {\bibfnamefont {E.}~\bibnamefont
  {Fradkin}}, \bibinfo {author} {\bibfnamefont {S.~A.}\ \bibnamefont
  {Kivelson}},\ and\ \bibinfo {author} {\bibfnamefont {J.~M.}\ \bibnamefont
  {Tranquada}},\ }\bibfield  {title} {\bibinfo {title} {{Colloquium: Theory of
  intertwined orders in high temperature superconductors}},\ }\href@noop {}
  {\bibfield  {journal} {\bibinfo  {journal} {Rev. Mod. Phys.}\ }\textbf
  {\bibinfo {volume} {87}},\ \bibinfo {pages} {457} (\bibinfo {year}
  {2015})}\BibitemShut {NoStop}%
\bibitem [{\citenamefont {Keimer}\ \emph {et~al.}(2015)\citenamefont {Keimer},
  \citenamefont {Kivelson}, \citenamefont {Norman}, \citenamefont {Uchida},\
  and\ \citenamefont {Zaanen}}]{Keimer2015}%
  \BibitemOpen
  \bibfield  {author} {\bibinfo {author} {\bibfnamefont {B.}~\bibnamefont
  {Keimer}}, \bibinfo {author} {\bibfnamefont {S.~A.}\ \bibnamefont
  {Kivelson}}, \bibinfo {author} {\bibfnamefont {M.~R.}\ \bibnamefont
  {Norman}}, \bibinfo {author} {\bibfnamefont {S.}~\bibnamefont {Uchida}},\
  and\ \bibinfo {author} {\bibfnamefont {J.}~\bibnamefont {Zaanen}},\
  }\bibfield  {title} {\bibinfo {title} {From quantum matter to
  high-temperature superconductivity in copper oxides},\ }\href@noop {}
  {\bibfield  {journal} {\bibinfo  {journal} {Nature}\ }\textbf {\bibinfo
  {volume} {518}},\ \bibinfo {pages} {179} (\bibinfo {year}
  {2015})}\BibitemShut {NoStop}%
\bibitem [{\citenamefont {Kivelson}\ \emph {et~al.}(2003)\citenamefont
  {Kivelson}, \citenamefont {Bindloss}, \citenamefont {Fradkin}, \citenamefont
  {Oganesyan}, \citenamefont {Tranquada}, \citenamefont {Kapitulnik},\ and\
  \citenamefont {Howald}}]{Kivelson03}%
  \BibitemOpen
  \bibfield  {author} {\bibinfo {author} {\bibfnamefont {S.~A.}\ \bibnamefont
  {Kivelson}}, \bibinfo {author} {\bibfnamefont {I.~P.}\ \bibnamefont
  {Bindloss}}, \bibinfo {author} {\bibfnamefont {E.}~\bibnamefont {Fradkin}},
  \bibinfo {author} {\bibfnamefont {V.}~\bibnamefont {Oganesyan}}, \bibinfo
  {author} {\bibfnamefont {J.~M.}\ \bibnamefont {Tranquada}}, \bibinfo {author}
  {\bibfnamefont {A.}~\bibnamefont {Kapitulnik}},\ and\ \bibinfo {author}
  {\bibfnamefont {C.}~\bibnamefont {Howald}},\ }\bibfield  {title} {\bibinfo
  {title} {How to detect fluctuating stripes in the high-temperature
  superconductors},\ }\href {https://doi.org/10.1103/RevModPhys.75.1201}
  {\bibfield  {journal} {\bibinfo  {journal} {Rev. Mod. Phys.}\ }\textbf
  {\bibinfo {volume} {75}},\ \bibinfo {pages} {1201} (\bibinfo {year}
  {2003})}\BibitemShut {NoStop}%
\bibitem [{\citenamefont {Arumugam}\ \emph {et~al.}(2000)\citenamefont
  {Arumugam}, \citenamefont {Mori}, \citenamefont {Takeshita}, \citenamefont
  {Takashima}, \citenamefont {Noda}, \citenamefont {Eisaki},\ and\
  \citenamefont {Uchida}}]{ARUMUGAM2000}%
  \BibitemOpen
  \bibfield  {author} {\bibinfo {author} {\bibfnamefont {S.}~\bibnamefont
  {Arumugam}}, \bibinfo {author} {\bibfnamefont {N.}~\bibnamefont {Mori}},
  \bibinfo {author} {\bibfnamefont {N.}~\bibnamefont {Takeshita}}, \bibinfo
  {author} {\bibfnamefont {H.}~\bibnamefont {Takashima}}, \bibinfo {author}
  {\bibfnamefont {T.}~\bibnamefont {Noda}}, \bibinfo {author} {\bibfnamefont
  {H.}~\bibnamefont {Eisaki}},\ and\ \bibinfo {author} {\bibfnamefont
  {S.}~\bibnamefont {Uchida}},\ }\bibfield  {title} {\bibinfo {title}
  {Transport measurements of {La-Nd-Sr-Cu-O} superconductors under hydrostatic
  and uniaxial pressure},\ }\href
  {https://doi.org/https://doi.org/10.1016/S0921-4534(00)01006-6} {\bibfield
  {journal} {\bibinfo  {journal} {Physica C: Superconductivity}\ }\textbf
  {\bibinfo {volume} {341-348}},\ \bibinfo {pages} {1759} (\bibinfo {year}
  {2000})}\BibitemShut {NoStop}%
\bibitem [{\citenamefont {Takeshita}\ \emph {et~al.}(2004)\citenamefont
  {Takeshita}, \citenamefont {Sasagawa}, \citenamefont {Sugioka}, \citenamefont
  {Tokura},\ and\ \citenamefont {Takagi}}]{Takeshita04}%
  \BibitemOpen
  \bibfield  {author} {\bibinfo {author} {\bibfnamefont {N.}~\bibnamefont
  {Takeshita}}, \bibinfo {author} {\bibfnamefont {T.}~\bibnamefont {Sasagawa}},
  \bibinfo {author} {\bibfnamefont {T.}~\bibnamefont {Sugioka}}, \bibinfo
  {author} {\bibfnamefont {Y.}~\bibnamefont {Tokura}},\ and\ \bibinfo {author}
  {\bibfnamefont {H.}~\bibnamefont {Takagi}},\ }\bibfield  {title} {\bibinfo
  {title} {Gigantic anisotropic uniaxial pressure effect on superconductivity
  within the {CuO$_2$} plane of {La$_{1.64}$Eu$_{0.2}$Sr$_{0.16}$CuO$_4$}:
  Strain control of stripe criticality},\ }\href
  {https://doi.org/10.1143/JPSJ.73.1123} {\bibfield  {journal} {\bibinfo
  {journal} {Journal of the Physical Society of Japan}\ }\textbf {\bibinfo
  {volume} {73}},\ \bibinfo {pages} {1123} (\bibinfo {year}
  {2004})}\BibitemShut {NoStop}%
\bibitem [{\citenamefont {Sasagawa}\ \emph {et~al.}(2005)\citenamefont
  {Sasagawa}, \citenamefont {Suryadijaya}, \citenamefont {Shimatani},\ and\
  \citenamefont {Takagi}}]{Sasagawa05}%
  \BibitemOpen
  \bibfield  {author} {\bibinfo {author} {\bibfnamefont {T.}~\bibnamefont
  {Sasagawa}}, \bibinfo {author} {\bibnamefont {Suryadijaya}}, \bibinfo
  {author} {\bibfnamefont {K.}~\bibnamefont {Shimatani}},\ and\ \bibinfo
  {author} {\bibfnamefont {H.}~\bibnamefont {Takagi}},\ }\bibfield  {title}
  {\bibinfo {title} {Control of stripes/superconductivity competition in
  {$\mathrm{(La,Eu,Sr)_2CuO_4}$} crystals using uniaxial pressure},\ }\href
  {https://doi.org/https://doi.org/10.1016/j.physb.2005.01.089} {\bibfield
  {journal} {\bibinfo  {journal} {Physica B: Condensed Matter}\ }\textbf
  {\bibinfo {volume} {359-361}},\ \bibinfo {pages} {436} (\bibinfo {year}
  {2005})}\BibitemShut {NoStop}%
\bibitem [{\citenamefont {Guguchia}\ \emph {et~al.}(2020)\citenamefont
  {Guguchia}, \citenamefont {Das}, \citenamefont {Wang}, \citenamefont
  {Adachi}, \citenamefont {Kitajima}, \citenamefont {Elender}, \citenamefont
  {Br\"uckner}, \citenamefont {Ghosh}, \citenamefont {Grinenko}, \citenamefont
  {Shiroka}, \citenamefont {M\"uller}, \citenamefont {Mudry}, \citenamefont
  {Baines}, \citenamefont {Bartkowiak}, \citenamefont {Koike}, \citenamefont
  {Amato}, \citenamefont {Tranquada}, \citenamefont {Klauss}, \citenamefont
  {Hicks},\ and\ \citenamefont {Luetkens}}]{Guguchia20}%
  \BibitemOpen
  \bibfield  {author} {\bibinfo {author} {\bibfnamefont {Z.}~\bibnamefont
  {Guguchia}}, \bibinfo {author} {\bibfnamefont {D.}~\bibnamefont {Das}},
  \bibinfo {author} {\bibfnamefont {C.~N.}\ \bibnamefont {Wang}}, \bibinfo
  {author} {\bibfnamefont {T.}~\bibnamefont {Adachi}}, \bibinfo {author}
  {\bibfnamefont {N.}~\bibnamefont {Kitajima}}, \bibinfo {author}
  {\bibfnamefont {M.}~\bibnamefont {Elender}}, \bibinfo {author} {\bibfnamefont
  {F.}~\bibnamefont {Br\"uckner}}, \bibinfo {author} {\bibfnamefont
  {S.}~\bibnamefont {Ghosh}}, \bibinfo {author} {\bibfnamefont
  {V.}~\bibnamefont {Grinenko}}, \bibinfo {author} {\bibfnamefont
  {T.}~\bibnamefont {Shiroka}}, \bibinfo {author} {\bibfnamefont
  {M.}~\bibnamefont {M\"uller}}, \bibinfo {author} {\bibfnamefont
  {C.}~\bibnamefont {Mudry}}, \bibinfo {author} {\bibfnamefont
  {C.}~\bibnamefont {Baines}}, \bibinfo {author} {\bibfnamefont
  {M.}~\bibnamefont {Bartkowiak}}, \bibinfo {author} {\bibfnamefont
  {Y.}~\bibnamefont {Koike}}, \bibinfo {author} {\bibfnamefont
  {A.}~\bibnamefont {Amato}}, \bibinfo {author} {\bibfnamefont {J.~M.}\
  \bibnamefont {Tranquada}}, \bibinfo {author} {\bibfnamefont {H.-H.}\
  \bibnamefont {Klauss}}, \bibinfo {author} {\bibfnamefont {C.~W.}\
  \bibnamefont {Hicks}},\ and\ \bibinfo {author} {\bibfnamefont
  {H.}~\bibnamefont {Luetkens}},\ }\bibfield  {title} {\bibinfo {title} {Using
  uniaxial stress to probe the relationship between competing superconducting
  states in a cuprate with spin-stripe order},\ }\href
  {https://doi.org/10.1103/PhysRevLett.125.097005} {\bibfield  {journal}
  {\bibinfo  {journal} {Phys. Rev. Lett.}\ }\textbf {\bibinfo {volume} {125}},\
  \bibinfo {pages} {097005} (\bibinfo {year} {2020})}\BibitemShut {NoStop}%
\bibitem [{\citenamefont {Kim}\ \emph {et~al.}(2018)\citenamefont {Kim},
  \citenamefont {Souliou}, \citenamefont {Barber}, \citenamefont {Lefrançois},
  \citenamefont {Minola}, \citenamefont {Tortora}, \citenamefont {Heid},
  \citenamefont {Nandi}, \citenamefont {Borzi}, \citenamefont {Garbarino},
  \citenamefont {Bosak}, \citenamefont {Porras}, \citenamefont {Loew},
  \citenamefont {König}, \citenamefont {Moll}, \citenamefont {Mackenzie},
  \citenamefont {Keimer}, \citenamefont {Hicks},\ and\ \citenamefont
  {Tacon}}]{KIm18}%
  \BibitemOpen
  \bibfield  {author} {\bibinfo {author} {\bibfnamefont {H.-H.}\ \bibnamefont
  {Kim}}, \bibinfo {author} {\bibfnamefont {S.~M.}\ \bibnamefont {Souliou}},
  \bibinfo {author} {\bibfnamefont {M.~E.}\ \bibnamefont {Barber}}, \bibinfo
  {author} {\bibfnamefont {E.}~\bibnamefont {Lefrançois}}, \bibinfo {author}
  {\bibfnamefont {M.}~\bibnamefont {Minola}}, \bibinfo {author} {\bibfnamefont
  {M.}~\bibnamefont {Tortora}}, \bibinfo {author} {\bibfnamefont
  {R.}~\bibnamefont {Heid}}, \bibinfo {author} {\bibfnamefont {N.}~\bibnamefont
  {Nandi}}, \bibinfo {author} {\bibfnamefont {R.~A.}\ \bibnamefont {Borzi}},
  \bibinfo {author} {\bibfnamefont {G.}~\bibnamefont {Garbarino}}, \bibinfo
  {author} {\bibfnamefont {A.}~\bibnamefont {Bosak}}, \bibinfo {author}
  {\bibfnamefont {J.}~\bibnamefont {Porras}}, \bibinfo {author} {\bibfnamefont
  {T.}~\bibnamefont {Loew}}, \bibinfo {author} {\bibfnamefont {M.}~\bibnamefont
  {König}}, \bibinfo {author} {\bibfnamefont {P.~J.~W.}\ \bibnamefont {Moll}},
  \bibinfo {author} {\bibfnamefont {A.~P.}\ \bibnamefont {Mackenzie}}, \bibinfo
  {author} {\bibfnamefont {B.}~\bibnamefont {Keimer}}, \bibinfo {author}
  {\bibfnamefont {C.~W.}\ \bibnamefont {Hicks}},\ and\ \bibinfo {author}
  {\bibfnamefont {M.~L.}\ \bibnamefont {Tacon}},\ }\bibfield  {title} {\bibinfo
  {title} {Uniaxial pressure control of competing orders in a high-temperature
  superconductor},\ }\href {https://doi.org/10.1126/science.aat4708} {\bibfield
   {journal} {\bibinfo  {journal} {Science}\ }\textbf {\bibinfo {volume}
  {362}},\ \bibinfo {pages} {1040} (\bibinfo {year} {2018})}\BibitemShut
  {NoStop}%
\bibitem [{\citenamefont {Kim}\ \emph {et~al.}(2021)\citenamefont {Kim},
  \citenamefont {Lefran\ifmmode~\mbox{\c{c}}\else \c{c}\fi{}ois}, \citenamefont
  {Kummer}, \citenamefont {Fumagalli}, \citenamefont {Brookes}, \citenamefont
  {Betto}, \citenamefont {Nakata}, \citenamefont {Tortora}, \citenamefont
  {Porras}, \citenamefont {Loew}, \citenamefont {Barber}, \citenamefont
  {Braicovich}, \citenamefont {Mackenzie}, \citenamefont {Hicks}, \citenamefont
  {Keimer}, \citenamefont {Minola},\ and\ \citenamefont {Le~Tacon}}]{Kim21}%
  \BibitemOpen
  \bibfield  {author} {\bibinfo {author} {\bibfnamefont {H.-H.}\ \bibnamefont
  {Kim}}, \bibinfo {author} {\bibfnamefont {E.}~\bibnamefont
  {Lefran\ifmmode~\mbox{\c{c}}\else \c{c}\fi{}ois}}, \bibinfo {author}
  {\bibfnamefont {K.}~\bibnamefont {Kummer}}, \bibinfo {author} {\bibfnamefont
  {R.}~\bibnamefont {Fumagalli}}, \bibinfo {author} {\bibfnamefont {N.~B.}\
  \bibnamefont {Brookes}}, \bibinfo {author} {\bibfnamefont {D.}~\bibnamefont
  {Betto}}, \bibinfo {author} {\bibfnamefont {S.}~\bibnamefont {Nakata}},
  \bibinfo {author} {\bibfnamefont {M.}~\bibnamefont {Tortora}}, \bibinfo
  {author} {\bibfnamefont {J.}~\bibnamefont {Porras}}, \bibinfo {author}
  {\bibfnamefont {T.}~\bibnamefont {Loew}}, \bibinfo {author} {\bibfnamefont
  {M.~E.}\ \bibnamefont {Barber}}, \bibinfo {author} {\bibfnamefont
  {L.}~\bibnamefont {Braicovich}}, \bibinfo {author} {\bibfnamefont {A.~P.}\
  \bibnamefont {Mackenzie}}, \bibinfo {author} {\bibfnamefont {C.~W.}\
  \bibnamefont {Hicks}}, \bibinfo {author} {\bibfnamefont {B.}~\bibnamefont
  {Keimer}}, \bibinfo {author} {\bibfnamefont {M.}~\bibnamefont {Minola}},\
  and\ \bibinfo {author} {\bibfnamefont {M.}~\bibnamefont {Le~Tacon}},\
  }\bibfield  {title} {\bibinfo {title} {Charge density waves in
  {${\mathrm{YBa}}_{2}{\mathrm{Cu}}_{3}{\mathrm{O}}_{6.67}$} probed by resonant
  x-ray scattering under uniaxial compression},\ }\href
  {https://doi.org/10.1103/PhysRevLett.126.037002} {\bibfield  {journal}
  {\bibinfo  {journal} {Phys. Rev. Lett.}\ }\textbf {\bibinfo {volume} {126}},\
  \bibinfo {pages} {037002} (\bibinfo {year} {2021})}\BibitemShut {NoStop}%
\bibitem [{\citenamefont {Choi}\ \emph {et~al.}(2022)\citenamefont {Choi},
  \citenamefont {Wang}, \citenamefont {J\"ohr}, \citenamefont {Christensen},
  \citenamefont {K\"uspert}, \citenamefont {Bucher}, \citenamefont {Biscette},
  \citenamefont {Fischer}, \citenamefont {H\"ucker}, \citenamefont {Kurosawa},
  \citenamefont {Momono}, \citenamefont {Oda}, \citenamefont {Ivashko},
  \citenamefont {Zimmermann}, \citenamefont {Janoschek},\ and\ \citenamefont
  {Chang}}]{Choi22}%
  \BibitemOpen
  \bibfield  {author} {\bibinfo {author} {\bibfnamefont {J.}~\bibnamefont
  {Choi}}, \bibinfo {author} {\bibfnamefont {Q.}~\bibnamefont {Wang}}, \bibinfo
  {author} {\bibfnamefont {S.}~\bibnamefont {J\"ohr}}, \bibinfo {author}
  {\bibfnamefont {N.~B.}\ \bibnamefont {Christensen}}, \bibinfo {author}
  {\bibfnamefont {J.}~\bibnamefont {K\"uspert}}, \bibinfo {author}
  {\bibfnamefont {D.}~\bibnamefont {Bucher}}, \bibinfo {author} {\bibfnamefont
  {D.}~\bibnamefont {Biscette}}, \bibinfo {author} {\bibfnamefont {M.~H.}\
  \bibnamefont {Fischer}}, \bibinfo {author} {\bibfnamefont {M.}~\bibnamefont
  {H\"ucker}}, \bibinfo {author} {\bibfnamefont {T.}~\bibnamefont {Kurosawa}},
  \bibinfo {author} {\bibfnamefont {N.}~\bibnamefont {Momono}}, \bibinfo
  {author} {\bibfnamefont {M.}~\bibnamefont {Oda}}, \bibinfo {author}
  {\bibfnamefont {O.}~\bibnamefont {Ivashko}}, \bibinfo {author} {\bibfnamefont
  {M.~v.}\ \bibnamefont {Zimmermann}}, \bibinfo {author} {\bibfnamefont
  {M.}~\bibnamefont {Janoschek}},\ and\ \bibinfo {author} {\bibfnamefont
  {J.}~\bibnamefont {Chang}},\ }\bibfield  {title} {\bibinfo {title} {Unveiling
  unequivocal charge stripe order in a prototypical cuprate superconductor},\
  }\href {https://doi.org/10.1103/PhysRevLett.128.207002} {\bibfield  {journal}
  {\bibinfo  {journal} {Phys. Rev. Lett.}\ }\textbf {\bibinfo {volume} {128}},\
  \bibinfo {pages} {207002} (\bibinfo {year} {2022})}\BibitemShut {NoStop}%
\bibitem [{\citenamefont {Boyle}\ \emph {et~al.}(2021)\citenamefont {Boyle},
  \citenamefont {Walker}, \citenamefont {Ruiz}, \citenamefont {Schierle},
  \citenamefont {Zhao}, \citenamefont {Boschini}, \citenamefont {Sutarto},
  \citenamefont {Boyko}, \citenamefont {Moore}, \citenamefont {Tamura},
  \citenamefont {He}, \citenamefont {Weschke}, \citenamefont {Gozar},
  \citenamefont {Peng}, \citenamefont {Komarek}, \citenamefont {Damascelli},
  \citenamefont {Sch\"u\ss{}ler-Langeheine}, \citenamefont {Frano},
  \citenamefont {da~Silva~Neto},\ and\ \citenamefont
  {Blanco-Canosa}}]{Boyle2021}%
  \BibitemOpen
  \bibfield  {author} {\bibinfo {author} {\bibfnamefont {T.~J.}\ \bibnamefont
  {Boyle}}, \bibinfo {author} {\bibfnamefont {M.}~\bibnamefont {Walker}},
  \bibinfo {author} {\bibfnamefont {A.}~\bibnamefont {Ruiz}}, \bibinfo {author}
  {\bibfnamefont {E.}~\bibnamefont {Schierle}}, \bibinfo {author}
  {\bibfnamefont {Z.}~\bibnamefont {Zhao}}, \bibinfo {author} {\bibfnamefont
  {F.}~\bibnamefont {Boschini}}, \bibinfo {author} {\bibfnamefont
  {R.}~\bibnamefont {Sutarto}}, \bibinfo {author} {\bibfnamefont {T.~D.}\
  \bibnamefont {Boyko}}, \bibinfo {author} {\bibfnamefont {W.}~\bibnamefont
  {Moore}}, \bibinfo {author} {\bibfnamefont {N.}~\bibnamefont {Tamura}},
  \bibinfo {author} {\bibfnamefont {F.}~\bibnamefont {He}}, \bibinfo {author}
  {\bibfnamefont {E.}~\bibnamefont {Weschke}}, \bibinfo {author} {\bibfnamefont
  {A.}~\bibnamefont {Gozar}}, \bibinfo {author} {\bibfnamefont
  {W.}~\bibnamefont {Peng}}, \bibinfo {author} {\bibfnamefont {A.~C.}\
  \bibnamefont {Komarek}}, \bibinfo {author} {\bibfnamefont {A.}~\bibnamefont
  {Damascelli}}, \bibinfo {author} {\bibfnamefont {C.}~\bibnamefont
  {Sch\"u\ss{}ler-Langeheine}}, \bibinfo {author} {\bibfnamefont
  {A.}~\bibnamefont {Frano}}, \bibinfo {author} {\bibfnamefont {E.~H.}\
  \bibnamefont {da~Silva~Neto}},\ and\ \bibinfo {author} {\bibfnamefont
  {S.}~\bibnamefont {Blanco-Canosa}},\ }\bibfield  {title} {\bibinfo {title}
  {Large response of charge stripes to uniaxial stress in
  {${\mathrm{La}}_{1.475}{\mathrm{Nd}}_{0.4}{\mathrm{Sr}}_{0.125}{\mathrm{CuO}}_{4}$}},\
  }\href {https://doi.org/10.1103/PhysRevResearch.3.L022004} {\bibfield
  {journal} {\bibinfo  {journal} {Phys. Rev. Research}\ }\textbf {\bibinfo
  {volume} {3}},\ \bibinfo {pages} {L022004} (\bibinfo {year}
  {2021})}\BibitemShut {NoStop}%
\bibitem [{\citenamefont {Wahlberg}\ \emph {et~al.}(2021)\citenamefont
  {Wahlberg}, \citenamefont {Arpaia}, \citenamefont {Seibold}, \citenamefont
  {Rossi}, \citenamefont {Fumagalli}, \citenamefont {Trabaldo}, \citenamefont
  {Brookes}, \citenamefont {Braicovich}, \citenamefont {Caprara}, \citenamefont
  {Gran}, \citenamefont {Ghiringhelli}, \citenamefont {Bauch},\ and\
  \citenamefont {Lombardi}}]{Wahlberg21}%
  \BibitemOpen
  \bibfield  {author} {\bibinfo {author} {\bibfnamefont {E.}~\bibnamefont
  {Wahlberg}}, \bibinfo {author} {\bibfnamefont {R.}~\bibnamefont {Arpaia}},
  \bibinfo {author} {\bibfnamefont {G.}~\bibnamefont {Seibold}}, \bibinfo
  {author} {\bibfnamefont {M.}~\bibnamefont {Rossi}}, \bibinfo {author}
  {\bibfnamefont {R.}~\bibnamefont {Fumagalli}}, \bibinfo {author}
  {\bibfnamefont {E.}~\bibnamefont {Trabaldo}}, \bibinfo {author}
  {\bibfnamefont {N.~B.}\ \bibnamefont {Brookes}}, \bibinfo {author}
  {\bibfnamefont {L.}~\bibnamefont {Braicovich}}, \bibinfo {author}
  {\bibfnamefont {S.}~\bibnamefont {Caprara}}, \bibinfo {author} {\bibfnamefont
  {U.}~\bibnamefont {Gran}}, \bibinfo {author} {\bibfnamefont {G.}~\bibnamefont
  {Ghiringhelli}}, \bibinfo {author} {\bibfnamefont {T.}~\bibnamefont
  {Bauch}},\ and\ \bibinfo {author} {\bibfnamefont {F.}~\bibnamefont
  {Lombardi}},\ }\bibfield  {title} {\bibinfo {title} {Restored strange metal
  phase through suppression of charge density waves in underdoped
  {YBa$_2$Cu$_3$O$_{7-\delta}$}},\ }\href@noop {} {\bibfield  {journal}
  {\bibinfo  {journal} {Science}\ }\textbf {\bibinfo {volume} {373}},\ \bibinfo
  {pages} {1506} (\bibinfo {year} {2021})}\BibitemShut {NoStop}%
\bibitem [{\citenamefont {Guguchia}\ \emph {et~al.}(2023)\citenamefont
  {Guguchia}, \citenamefont {Das}, \citenamefont {Simutis}, \citenamefont
  {Adachi}, \citenamefont {K{\"u}spert}, \citenamefont {Kitajima},
  \citenamefont {Elender}, \citenamefont {Grinenko}, \citenamefont {Ivashko},
  \citenamefont {Zimmermann} \emph {et~al.}}]{guguchia2023designing}%
  \BibitemOpen
  \bibfield  {author} {\bibinfo {author} {\bibfnamefont {Z.}~\bibnamefont
  {Guguchia}}, \bibinfo {author} {\bibfnamefont {D.}~\bibnamefont {Das}},
  \bibinfo {author} {\bibfnamefont {G.}~\bibnamefont {Simutis}}, \bibinfo
  {author} {\bibfnamefont {T.}~\bibnamefont {Adachi}}, \bibinfo {author}
  {\bibfnamefont {J.}~\bibnamefont {K{\"u}spert}}, \bibinfo {author}
  {\bibfnamefont {N.}~\bibnamefont {Kitajima}}, \bibinfo {author}
  {\bibfnamefont {M.}~\bibnamefont {Elender}}, \bibinfo {author} {\bibfnamefont
  {V.}~\bibnamefont {Grinenko}}, \bibinfo {author} {\bibfnamefont
  {O.}~\bibnamefont {Ivashko}}, \bibinfo {author} {\bibfnamefont {M.~v.}\
  \bibnamefont {Zimmermann}}, \emph {et~al.},\ }\bibfield  {title} {\bibinfo
  {title} {Designing the stripe-ordered cuprate phase diagram through
  uniaxial-stress},\ }\href@noop {} {\bibfield  {journal} {\bibinfo  {journal}
  {arXiv preprint arXiv:2302.07015}\ } (\bibinfo {year} {2023})}\BibitemShut
  {NoStop}%
\bibitem [{\citenamefont {Wang}\ \emph {et~al.}(2022)\citenamefont {Wang},
  \citenamefont {von Arx}, \citenamefont {Mazzone}, \citenamefont {Mustafi},
  \citenamefont {Horio}, \citenamefont {K{\"u}spert}, \citenamefont {Choi},
  \citenamefont {Bucher}, \citenamefont {Wo}, \citenamefont {Zhao},
  \citenamefont {Zhang}, \citenamefont {Asmara}, \citenamefont {Sassa},
  \citenamefont {M{\aa}nsson}, \citenamefont {Christensen}, \citenamefont
  {Janoschek}, \citenamefont {Kurosawa}, \citenamefont {Momono}, \citenamefont
  {Oda}, \citenamefont {Fischer}, \citenamefont {Schmitt},\ and\ \citenamefont
  {Chang}}]{Wang22}%
  \BibitemOpen
  \bibfield  {author} {\bibinfo {author} {\bibfnamefont {Q.}~\bibnamefont
  {Wang}}, \bibinfo {author} {\bibfnamefont {K.}~\bibnamefont {von Arx}},
  \bibinfo {author} {\bibfnamefont {D.~G.}\ \bibnamefont {Mazzone}}, \bibinfo
  {author} {\bibfnamefont {S.}~\bibnamefont {Mustafi}}, \bibinfo {author}
  {\bibfnamefont {M.}~\bibnamefont {Horio}}, \bibinfo {author} {\bibfnamefont
  {J.}~\bibnamefont {K{\"u}spert}}, \bibinfo {author} {\bibfnamefont
  {J.}~\bibnamefont {Choi}}, \bibinfo {author} {\bibfnamefont {D.}~\bibnamefont
  {Bucher}}, \bibinfo {author} {\bibfnamefont {H.}~\bibnamefont {Wo}}, \bibinfo
  {author} {\bibfnamefont {J.}~\bibnamefont {Zhao}}, \bibinfo {author}
  {\bibfnamefont {W.}~\bibnamefont {Zhang}}, \bibinfo {author} {\bibfnamefont
  {T.~C.}\ \bibnamefont {Asmara}}, \bibinfo {author} {\bibfnamefont
  {Y.}~\bibnamefont {Sassa}}, \bibinfo {author} {\bibfnamefont
  {M.}~\bibnamefont {M{\aa}nsson}}, \bibinfo {author} {\bibfnamefont {N.~B.}\
  \bibnamefont {Christensen}}, \bibinfo {author} {\bibfnamefont
  {M.}~\bibnamefont {Janoschek}}, \bibinfo {author} {\bibfnamefont
  {T.}~\bibnamefont {Kurosawa}}, \bibinfo {author} {\bibfnamefont
  {N.}~\bibnamefont {Momono}}, \bibinfo {author} {\bibfnamefont
  {M.}~\bibnamefont {Oda}}, \bibinfo {author} {\bibfnamefont {M.~H.}\
  \bibnamefont {Fischer}}, \bibinfo {author} {\bibfnamefont {T.}~\bibnamefont
  {Schmitt}},\ and\ \bibinfo {author} {\bibfnamefont {J.}~\bibnamefont
  {Chang}},\ }\bibfield  {title} {\bibinfo {title} {{Uniaxial pressure induced
  stripe order rotation in {La$_{1.88}$Sr$_{0.12}$CuO$_4$}}},\ }\href@noop {}
  {\bibfield  {journal} {\bibinfo  {journal} {Nature Communications}\ }\textbf
  {\bibinfo {volume} {13}},\ \bibinfo {pages} {1795} (\bibinfo {year}
  {2022})}\BibitemShut {NoStop}%
\bibitem [{\citenamefont {Frison}\ \emph {et~al.}(2022)\citenamefont {Frison},
  \citenamefont {K\"uspert}, \citenamefont {Wang}, \citenamefont {Ivashko},
  \citenamefont {Zimmermann}, \citenamefont {Meven}, \citenamefont {Bucher},
  \citenamefont {Larsen}, \citenamefont {Niedermayer}, \citenamefont
  {Janoschek}, \citenamefont {Kurosawa}, \citenamefont {Momono}, \citenamefont
  {Oda}, \citenamefont {Christensen},\ and\ \citenamefont
  {Chang}}]{LSCO_Frison}%
  \BibitemOpen
  \bibfield  {author} {\bibinfo {author} {\bibfnamefont {R.}~\bibnamefont
  {Frison}}, \bibinfo {author} {\bibfnamefont {J.}~\bibnamefont {K\"uspert}},
  \bibinfo {author} {\bibfnamefont {Q.}~\bibnamefont {Wang}}, \bibinfo {author}
  {\bibfnamefont {O.}~\bibnamefont {Ivashko}}, \bibinfo {author} {\bibfnamefont
  {M.~v.}\ \bibnamefont {Zimmermann}}, \bibinfo {author} {\bibfnamefont
  {M.}~\bibnamefont {Meven}}, \bibinfo {author} {\bibfnamefont
  {D.}~\bibnamefont {Bucher}}, \bibinfo {author} {\bibfnamefont
  {J.}~\bibnamefont {Larsen}}, \bibinfo {author} {\bibfnamefont
  {C.}~\bibnamefont {Niedermayer}}, \bibinfo {author} {\bibfnamefont
  {M.}~\bibnamefont {Janoschek}}, \bibinfo {author} {\bibfnamefont
  {T.}~\bibnamefont {Kurosawa}}, \bibinfo {author} {\bibfnamefont
  {N.}~\bibnamefont {Momono}}, \bibinfo {author} {\bibfnamefont
  {M.}~\bibnamefont {Oda}}, \bibinfo {author} {\bibfnamefont {N.~B.}\
  \bibnamefont {Christensen}},\ and\ \bibinfo {author} {\bibfnamefont
  {J.}~\bibnamefont {Chang}},\ }\bibfield  {title} {\bibinfo {title} {Crystal
  symmetry of stripe-ordered
  {${\mathrm{La}}_{1.88}{\mathrm{Sr}}_{0.12}{\mathrm{CuO}}_{4}$}},\ }\href
  {https://doi.org/10.1103/PhysRevB.105.224113} {\bibfield  {journal} {\bibinfo
   {journal} {Phys. Rev. B}\ }\textbf {\bibinfo {volume} {105}},\ \bibinfo
  {pages} {224113} (\bibinfo {year} {2022})}\BibitemShut {NoStop}%
\bibitem [{\citenamefont {Vaknin}\ \emph {et~al.}(1987)\citenamefont {Vaknin},
  \citenamefont {Sinha}, \citenamefont {Moncton}, \citenamefont {Johnston},
  \citenamefont {Newsam}, \citenamefont {Safinya},\ and\ \citenamefont
  {King}}]{LSCO_Vaknin}%
  \BibitemOpen
  \bibfield  {author} {\bibinfo {author} {\bibfnamefont {D.}~\bibnamefont
  {Vaknin}}, \bibinfo {author} {\bibfnamefont {S.~K.}\ \bibnamefont {Sinha}},
  \bibinfo {author} {\bibfnamefont {D.~E.}\ \bibnamefont {Moncton}}, \bibinfo
  {author} {\bibfnamefont {D.~C.}\ \bibnamefont {Johnston}}, \bibinfo {author}
  {\bibfnamefont {J.~M.}\ \bibnamefont {Newsam}}, \bibinfo {author}
  {\bibfnamefont {C.~R.}\ \bibnamefont {Safinya}},\ and\ \bibinfo {author}
  {\bibfnamefont {H.~E.}\ \bibnamefont {King}},\ }\bibfield  {title} {\bibinfo
  {title} {Antiferromagnetism in
  {${\mathrm{La}}_{2}$${\mathrm{CuO}}_{4\mathrm{\ensuremath{-}}\mathrm{y}}$}},\
  }\href {https://doi.org/10.1103/PhysRevLett.58.2802} {\bibfield  {journal}
  {\bibinfo  {journal} {Phys. Rev. Lett.}\ }\textbf {\bibinfo {volume} {58}},\
  \bibinfo {pages} {2802} (\bibinfo {year} {1987})}\BibitemShut {NoStop}%
\bibitem [{\citenamefont {Axe}\ \emph {et~al.}(1989)\citenamefont {Axe},
  \citenamefont {Moudden}, \citenamefont {Hohlwein}, \citenamefont {Cox},
  \citenamefont {Mohanty}, \citenamefont {Moodenbaugh},\ and\ \citenamefont
  {Xu}}]{Axe89}%
  \BibitemOpen
  \bibfield  {author} {\bibinfo {author} {\bibfnamefont {J.~D.}\ \bibnamefont
  {Axe}}, \bibinfo {author} {\bibfnamefont {A.~H.}\ \bibnamefont {Moudden}},
  \bibinfo {author} {\bibfnamefont {D.}~\bibnamefont {Hohlwein}}, \bibinfo
  {author} {\bibfnamefont {D.~E.}\ \bibnamefont {Cox}}, \bibinfo {author}
  {\bibfnamefont {K.~M.}\ \bibnamefont {Mohanty}}, \bibinfo {author}
  {\bibfnamefont {A.~R.}\ \bibnamefont {Moodenbaugh}},\ and\ \bibinfo {author}
  {\bibfnamefont {Y.}~\bibnamefont {Xu}},\ }\bibfield  {title} {\bibinfo
  {title} {Structural phase transformations and superconductivity in
  {${\mathrm{La}}_{2\mathrm{-}\mathrm{x}}$${\mathrm{Ba}}_{\mathrm{x}}$${\mathrm{CuO}}_{4}$}},\
  }\href {https://doi.org/10.1103/PhysRevLett.62.2751} {\bibfield  {journal}
  {\bibinfo  {journal} {Phys. Rev. Lett.}\ }\textbf {\bibinfo {volume} {62}},\
  \bibinfo {pages} {2751} (\bibinfo {year} {1989})}\BibitemShut {NoStop}%
\bibitem [{\citenamefont {Suzuki}\ and\ \citenamefont
  {Fujita}(1989)}]{Suzuki89}%
  \BibitemOpen
  \bibfield  {author} {\bibinfo {author} {\bibfnamefont {T.}~\bibnamefont
  {Suzuki}}\ and\ \bibinfo {author} {\bibfnamefont {T.}~\bibnamefont
  {Fujita}},\ }\bibfield  {title} {\bibinfo {title} {{Anomalous Change in
  Crystalline Structure of {(La$_{1-x}$Ba$_x$)$_2$CuO$_{4-\delta}$}}},\
  }\href@noop {} {\bibfield  {journal} {\bibinfo  {journal} {Journal of the
  Physical Society of Japan}\ }\textbf {\bibinfo {volume} {58}},\ \bibinfo
  {pages} {1883} (\bibinfo {year} {1989})}\BibitemShut {NoStop}%
\bibitem [{\citenamefont {H{\"u}cker}\ \emph {et~al.}(2011)\citenamefont
  {H{\"u}cker}, \citenamefont {v.~Zimmermann}, \citenamefont {Gu},
  \citenamefont {Xu}, \citenamefont {Wen}, \citenamefont {Xu}, \citenamefont
  {Kang}, \citenamefont {Zheludev},\ and\ \citenamefont
  {Tranquada}}]{Hucker11}%
  \BibitemOpen
  \bibfield  {author} {\bibinfo {author} {\bibfnamefont {M.}~\bibnamefont
  {H{\"u}cker}}, \bibinfo {author} {\bibfnamefont {M.}~\bibnamefont
  {v.~Zimmermann}}, \bibinfo {author} {\bibfnamefont {G.~D.}\ \bibnamefont
  {Gu}}, \bibinfo {author} {\bibfnamefont {Z.~J.}\ \bibnamefont {Xu}}, \bibinfo
  {author} {\bibfnamefont {J.~S.}\ \bibnamefont {Wen}}, \bibinfo {author}
  {\bibfnamefont {G.}~\bibnamefont {Xu}}, \bibinfo {author} {\bibfnamefont
  {H.~J.}\ \bibnamefont {Kang}}, \bibinfo {author} {\bibfnamefont
  {A.}~\bibnamefont {Zheludev}},\ and\ \bibinfo {author} {\bibfnamefont
  {J.~M.}\ \bibnamefont {Tranquada}},\ }\bibfield  {title} {\bibinfo {title}
  {Stripe order in superconducting {La$_{2-x}$Ba$_x$CuO$_4$ ($0.095\le x\le
  0.155$)}},\ }\href {https://doi.org/10.1103/PhysRevB.83.104506} {\bibfield
  {journal} {\bibinfo  {journal} {Phys. Rev. B}\ }\textbf {\bibinfo {volume}
  {83}},\ \bibinfo {pages} {104506} (\bibinfo {year} {2011})}\BibitemShut
  {NoStop}%
\bibitem [{\citenamefont {Hücker}(2012)}]{Hucker12}%
  \BibitemOpen
  \bibfield  {author} {\bibinfo {author} {\bibfnamefont {M.}~\bibnamefont
  {Hücker}},\ }\bibfield  {title} {\bibinfo {title} {Structural aspects of
  materials with static stripe order},\ }\href
  {https://doi.org/https://doi.org/10.1016/j.physc.2012.04.035} {\bibfield
  {journal} {\bibinfo  {journal} {Physica C: Superconductivity}\ }\textbf
  {\bibinfo {volume} {481}},\ \bibinfo {pages} {3} (\bibinfo {year} {2012})},\
  \bibinfo {note} {stripes and Electronic Liquid Crystals in Strongly
  Correlated Materials}\BibitemShut {NoStop}%
\bibitem [{\citenamefont {Yamase}\ and\ \citenamefont
  {Kohno}(2000)}]{Yamase2000}%
  \BibitemOpen
  \bibfield  {author} {\bibinfo {author} {\bibfnamefont {H.}~\bibnamefont
  {Yamase}}\ and\ \bibinfo {author} {\bibfnamefont {H.}~\bibnamefont {Kohno}},\
  }\bibfield  {title} {\bibinfo {title} {Instability toward formation of
  quasi-one-dimensional {F}ermi surface in two-dimensional {t-J} model},\
  }\href {https://doi.org/10.1143/JPSJ.69.2151} {\bibfield  {journal} {\bibinfo
   {journal} {Journal of the Physical Society of Japan}\ }\textbf {\bibinfo
  {volume} {69}},\ \bibinfo {pages} {2151} (\bibinfo {year}
  {2000})}\BibitemShut {NoStop}%
\bibitem [{\citenamefont {Kampf}\ \emph {et~al.}(2001)\citenamefont {Kampf},
  \citenamefont {Scalapino},\ and\ \citenamefont {White}}]{Kampf01}%
  \BibitemOpen
  \bibfield  {author} {\bibinfo {author} {\bibfnamefont {A.~P.}\ \bibnamefont
  {Kampf}}, \bibinfo {author} {\bibfnamefont {D.~J.}\ \bibnamefont
  {Scalapino}},\ and\ \bibinfo {author} {\bibfnamefont {S.~R.}\ \bibnamefont
  {White}},\ }\bibfield  {title} {\bibinfo {title} {Stripe orientation in an
  anisotropic $t-$\textit{J} model},\ }\href
  {https://doi.org/10.1103/PhysRevB.64.052509} {\bibfield  {journal} {\bibinfo
  {journal} {Phys. Rev. B}\ }\textbf {\bibinfo {volume} {64}},\ \bibinfo
  {pages} {052509} (\bibinfo {year} {2001})}\BibitemShut {NoStop}%
\bibitem [{\citenamefont {suppinfo}\ \citenamefont {suppinfo},
  \citenamefont {suppinfo},\ and\ \citenamefont {White}}]{suppinfo}%
  \BibitemOpen
  {\bibinfo {title} {See \href{https://journals.aps.org/prb/supplemental/10.1103/PhysRevB.108.L121113}{Supplemental Material} for a description of the uniaxial stress device; sample preparation and orientation; determination of limits on the strain percentage; and modeling of x-ray fluorescence background from the sample. The Supplemental Material also contains Refs. [38-43]}\ }\BibitemShut {NoStop}%
\bibitem [{\citenamefont {Sears}\ \emph {et~al.}(2023)\citenamefont {Sears},
  \citenamefont {Shen}, \citenamefont {Krogstad}, \citenamefont {Miao},
  \citenamefont {Bozin}, \citenamefont {Robinson}, \citenamefont {Gu},
  \citenamefont {Osborn}, \citenamefont {Rosenkranz}, \citenamefont
  {Tranquada},\ and\ \citenamefont {Dean}}]{Sears2023}%
  \BibitemOpen
  \bibfield  {author} {\bibinfo {author} {\bibfnamefont {J.}~\bibnamefont
  {Sears}}, \bibinfo {author} {\bibfnamefont {Y.}~\bibnamefont {Shen}},
  \bibinfo {author} {\bibfnamefont {M.~J.}\ \bibnamefont {Krogstad}}, \bibinfo
  {author} {\bibfnamefont {H.}~\bibnamefont {Miao}}, \bibinfo {author}
  {\bibfnamefont {E.~S.}\ \bibnamefont {Bozin}}, \bibinfo {author}
  {\bibfnamefont {I.~K.}\ \bibnamefont {Robinson}}, \bibinfo {author}
  {\bibfnamefont {G.~D.}\ \bibnamefont {Gu}}, \bibinfo {author} {\bibfnamefont
  {R.}~\bibnamefont {Osborn}}, \bibinfo {author} {\bibfnamefont
  {S.}~\bibnamefont {Rosenkranz}}, \bibinfo {author} {\bibfnamefont {J.~M.}\
  \bibnamefont {Tranquada}},\ and\ \bibinfo {author} {\bibfnamefont {M.~P.~M.}\
  \bibnamefont {Dean}},\ }\bibfield  {title} {\bibinfo {title} {Structure of
  charge density waves in
  {${\mathrm{La}}_{1.875}{\mathrm{Ba}}_{0.125}{\mathrm{CuO}}_{4}$}},\ }\href
{https://doi.org/10.1103/PhysRevB.107.115125} {\bibfield  {journal} {\bibinfo
   {journal} {Phys. Rev. B}\ }\textbf {\bibinfo {volume} {107}},\ \bibinfo
  {pages} {115125} (\bibinfo {year} {2023})}\BibitemShut {NoStop}%
\bibitem [{\citenamefont {Gupta}\ \emph {et~al.}(2021)\citenamefont {Gupta},
  \citenamefont {McMahon}, \citenamefont {Sutarto}, \citenamefont {Shi},
  \citenamefont {Gong}, \citenamefont {Wei}, \citenamefont {Shen},
  \citenamefont {He}, \citenamefont {Ma}, \citenamefont {Dragomir},
  \citenamefont {Gaulin},\ and\ \citenamefont {Hawthorn}}]{Gupta21}%
  \BibitemOpen
  \bibfield  {author} {\bibinfo {author} {\bibfnamefont {N.~K.}\ \bibnamefont
  {Gupta}}, \bibinfo {author} {\bibfnamefont {C.}~\bibnamefont {McMahon}},
  \bibinfo {author} {\bibfnamefont {R.}~\bibnamefont {Sutarto}}, \bibinfo
  {author} {\bibfnamefont {T.}~\bibnamefont {Shi}}, \bibinfo {author}
  {\bibfnamefont {R.}~\bibnamefont {Gong}}, \bibinfo {author} {\bibfnamefont
  {H.~I.}\ \bibnamefont {Wei}}, \bibinfo {author} {\bibfnamefont {K.~M.}\
  \bibnamefont {Shen}}, \bibinfo {author} {\bibfnamefont {F.}~\bibnamefont
  {He}}, \bibinfo {author} {\bibfnamefont {Q.}~\bibnamefont {Ma}}, \bibinfo
  {author} {\bibfnamefont {M.}~\bibnamefont {Dragomir}}, \bibinfo {author}
  {\bibfnamefont {B.~D.}\ \bibnamefont {Gaulin}},\ and\ \bibinfo {author}
  {\bibfnamefont {D.~G.}\ \bibnamefont {Hawthorn}},\ }\bibfield  {title}
  {\bibinfo {title} {Vanishing nematic order beyond the pseudogap phase in
  overdoped cuprate superconductors},\ }\href
  {https://doi.org/10.1073/pnas.2106881118} {\bibfield  {journal} {\bibinfo
  {journal} {Proceedings of the National Academy of Sciences}\ }\textbf
  {\bibinfo {volume} {118}},\ \bibinfo {pages} {e2106881118} (\bibinfo {year}
  {2021})}\BibitemShut {NoStop}%
\bibitem [{\citenamefont {Fink}\ \emph {et~al.}(2011)\citenamefont {Fink},
  \citenamefont {Soltwisch}, \citenamefont {Geck}, \citenamefont {Schierle},
  \citenamefont {Weschke},\ and\ \citenamefont {B{\"u}chner}}]{Fink11}%
  \BibitemOpen
  \bibfield  {author} {\bibinfo {author} {\bibfnamefont {J.}~\bibnamefont
  {Fink}}, \bibinfo {author} {\bibfnamefont {V.}~\bibnamefont {Soltwisch}},
  \bibinfo {author} {\bibfnamefont {J.}~\bibnamefont {Geck}}, \bibinfo {author}
  {\bibfnamefont {E.}~\bibnamefont {Schierle}}, \bibinfo {author}
  {\bibfnamefont {E.}~\bibnamefont {Weschke}},\ and\ \bibinfo {author}
  {\bibfnamefont {B.}~\bibnamefont {B{\"u}chner}},\ }\bibfield  {title}
  {\bibinfo {title} {Phase diagram of charge order in
  {La$_{1.8-x}$Eu$_{0.2}$Sr$_x$CuO$_4$} from resonant soft x-ray diffraction},\
  }\href {https://doi.org/10.1103/PhysRevB.83.092503} {\bibfield  {journal}
  {\bibinfo  {journal} {Phys. Rev. B}\ }\textbf {\bibinfo {volume} {83}},\
  \bibinfo {pages} {092503} (\bibinfo {year} {2011})}\BibitemShut {NoStop}%
\bibitem [{\citenamefont {Wilkins}\ \emph {et~al.}(2011)\citenamefont
  {Wilkins}, \citenamefont {Dean}, \citenamefont {Fink}, \citenamefont
  {H\"ucker}, \citenamefont {Geck}, \citenamefont {Soltwisch}, \citenamefont
  {Schierle}, \citenamefont {Weschke}, \citenamefont {Gu}, \citenamefont
  {Uchida}, \citenamefont {Ichikawa}, \citenamefont {Tranquada},\ and\
  \citenamefont {Hill}}]{Wilkins11}%
  \BibitemOpen
  \bibfield  {author} {\bibinfo {author} {\bibfnamefont {S.~B.}\ \bibnamefont
  {Wilkins}}, \bibinfo {author} {\bibfnamefont {M.~P.~M.}\ \bibnamefont
  {Dean}}, \bibinfo {author} {\bibfnamefont {J.}~\bibnamefont {Fink}}, \bibinfo
  {author} {\bibfnamefont {M.}~\bibnamefont {H\"ucker}}, \bibinfo {author}
  {\bibfnamefont {J.}~\bibnamefont {Geck}}, \bibinfo {author} {\bibfnamefont
  {V.}~\bibnamefont {Soltwisch}}, \bibinfo {author} {\bibfnamefont
  {E.}~\bibnamefont {Schierle}}, \bibinfo {author} {\bibfnamefont
  {E.}~\bibnamefont {Weschke}}, \bibinfo {author} {\bibfnamefont
  {G.}~\bibnamefont {Gu}}, \bibinfo {author} {\bibfnamefont {S.}~\bibnamefont
  {Uchida}}, \bibinfo {author} {\bibfnamefont {N.}~\bibnamefont {Ichikawa}},
  \bibinfo {author} {\bibfnamefont {J.~M.}\ \bibnamefont {Tranquada}},\ and\
  \bibinfo {author} {\bibfnamefont {J.~P.}\ \bibnamefont {Hill}},\ }\bibfield
  {title} {\bibinfo {title} {Comparison of stripe modulations in
  {La$_{1.875}$Ba$_{0.125}$CuO$_{4}$} and
  {La$_{1.48}$Nd$_{0.4}$Sr$_{0.12}$CuO$_{4}$}},\ }\href
  {https://doi.org/10.1103/PhysRevB.84.195101} {\bibfield  {journal} {\bibinfo
  {journal} {Phys. Rev. B}\ }\textbf {\bibinfo {volume} {84}},\ \bibinfo
  {pages} {195101} (\bibinfo {year} {2011})}\BibitemShut {NoStop}%
\bibitem [{\citenamefont {Arpaia}\ and\ \citenamefont
  {Ghiringhelli}(2021)}]{Arpaia21}%
  \BibitemOpen
  \bibfield  {author} {\bibinfo {author} {\bibfnamefont {R.}~\bibnamefont
  {Arpaia}}\ and\ \bibinfo {author} {\bibfnamefont {G.}~\bibnamefont
  {Ghiringhelli}},\ }\bibfield  {title} {\bibinfo {title} {Charge order at high
  temperature in cuprate superconductors},\ }\href
  {https://doi.org/10.7566/JPSJ.90.111005} {\bibfield  {journal} {\bibinfo
  {journal} {Journal of the Physical Society of Japan}\ }\textbf {\bibinfo
  {volume} {90}},\ \bibinfo {pages} {111005} (\bibinfo {year}
  {2021})}\BibitemShut {NoStop}\bibitem [{\citenamefont {Bluschke}\ \emph {et~al.}(2022)\citenamefont
  {Bluschke}, \citenamefont {Gupta}, \citenamefont {Jang}, \citenamefont
  {Husain}, \citenamefont {Lee}, \citenamefont {Na}, \citenamefont {Remedios},
  \citenamefont {Park}, \citenamefont {Kim}, \citenamefont {Jang},
  \citenamefont {Choi}, \citenamefont {Sutarto}, \citenamefont {Reid},
  \citenamefont {Dakovski}, \citenamefont {Coslovich}, \citenamefont {Nguyen},
  \citenamefont {Burdet}, \citenamefont {Lin}, \citenamefont {Revcolevschi},
  \citenamefont {Park}, \citenamefont {Geck}, \citenamefont {Turner},
  \citenamefont {Damascelli},\ and\ \citenamefont {Hawthorn}}]{bluschke2022}%
  \BibitemOpen
  \bibfield  {author} {\bibinfo {author} {\bibfnamefont {M.}~\bibnamefont
  {Bluschke}}, \bibinfo {author} {\bibfnamefont {N.~K.}\ \bibnamefont {Gupta}},
  \bibinfo {author} {\bibfnamefont {H.}~\bibnamefont {Jang}}, \bibinfo {author}
  {\bibfnamefont {A.~A.}\ \bibnamefont {Husain}}, \bibinfo {author}
  {\bibfnamefont {B.}~\bibnamefont {Lee}}, \bibinfo {author} {\bibfnamefont
  {M.}~\bibnamefont {Na}}, \bibinfo {author} {\bibfnamefont {B.~D.}\
  \bibnamefont {Remedios}}, \bibinfo {author} {\bibfnamefont {S.-Y.}\
  \bibnamefont {Park}}, \bibinfo {author} {\bibfnamefont {M.}~\bibnamefont
  {Kim}}, \bibinfo {author} {\bibfnamefont {D.}~\bibnamefont {Jang}}, \bibinfo
  {author} {\bibfnamefont {H.}~\bibnamefont {Choi}}, \bibinfo {author}
  {\bibfnamefont {R.}~\bibnamefont {Sutarto}}, \bibinfo {author} {\bibfnamefont
  {A.~H.}\ \bibnamefont {Reid}}, \bibinfo {author} {\bibfnamefont {G.~L.}\
  \bibnamefont {Dakovski}}, \bibinfo {author} {\bibfnamefont {G.}~\bibnamefont
  {Coslovich}}, \bibinfo {author} {\bibfnamefont {Q.~L.}\ \bibnamefont
  {Nguyen}}, \bibinfo {author} {\bibfnamefont {N.~G.}\ \bibnamefont {Burdet}},
  \bibinfo {author} {\bibfnamefont {M.-F.}\ \bibnamefont {Lin}}, \bibinfo
  {author} {\bibfnamefont {A.}~\bibnamefont {Revcolevschi}}, \bibinfo {author}
  {\bibfnamefont {J.-H.}\ \bibnamefont {Park}}, \bibinfo {author}
  {\bibfnamefont {J.}~\bibnamefont {Geck}}, \bibinfo {author} {\bibfnamefont
  {J.~J.}\ \bibnamefont {Turner}}, \bibinfo {author} {\bibfnamefont
  {A.}~\bibnamefont {Damascelli}},\ and\ \bibinfo {author} {\bibfnamefont
  {D.~G.}\ \bibnamefont {Hawthorn}},\ }\href@noop {} {\bibinfo {title}
  {Nematicity dynamics in the charge-density-wave phase of a cuprate
  superconductor}} (\bibinfo {year} {2022}),\ \Eprint
  {https://arxiv.org/abs/2209.11528} {arXiv:2209.11528 [cond-mat.str-el]}\BibitemShut {NoStop}%
\bibitem [{\citenamefont {Hawthorn}\ \emph {et~al.}(2011)\citenamefont
  {Hawthorn}, \citenamefont {He}, \citenamefont {Venema}, \citenamefont
  {Davis}, \citenamefont {Achkar}, \citenamefont {Zhang}, \citenamefont
  {Sutarto}, \citenamefont {Wadati}, \citenamefont {Radi}, \citenamefont
  {Wilson}, \citenamefont {Wright}, \citenamefont {Shen}, \citenamefont {Geck},
  \citenamefont {Zhang}, \citenamefont {Nov{\'a}k},\ and\ \citenamefont
  {Sawatzky}}]{Hawthorn11a}%
  \BibitemOpen
  \bibfield  {author} {\bibinfo {author} {\bibfnamefont {D.~G.}\ \bibnamefont
  {Hawthorn}}, \bibinfo {author} {\bibfnamefont {F.}~\bibnamefont {He}},
  \bibinfo {author} {\bibfnamefont {L.}~\bibnamefont {Venema}}, \bibinfo
  {author} {\bibfnamefont {H.}~\bibnamefont {Davis}}, \bibinfo {author}
  {\bibfnamefont {A.~J.}\ \bibnamefont {Achkar}}, \bibinfo {author}
  {\bibfnamefont {J.}~\bibnamefont {Zhang}}, \bibinfo {author} {\bibfnamefont
  {R.}~\bibnamefont {Sutarto}}, \bibinfo {author} {\bibfnamefont
  {H.}~\bibnamefont {Wadati}}, \bibinfo {author} {\bibfnamefont
  {A.}~\bibnamefont {Radi}}, \bibinfo {author} {\bibfnamefont {T.}~\bibnamefont
  {Wilson}}, \bibinfo {author} {\bibfnamefont {G.}~\bibnamefont {Wright}},
  \bibinfo {author} {\bibfnamefont {K.~M.}\ \bibnamefont {Shen}}, \bibinfo
  {author} {\bibfnamefont {J.}~\bibnamefont {Geck}}, \bibinfo {author}
  {\bibfnamefont {H.}~\bibnamefont {Zhang}}, \bibinfo {author} {\bibfnamefont
  {V.}~\bibnamefont {Nov{\'a}k}},\ and\ \bibinfo {author} {\bibfnamefont
  {G.~A.}\ \bibnamefont {Sawatzky}},\ }\bibfield  {title} {\bibinfo {title} {An
  in-vacuum diffractometer for resonant elastic soft x-ray scattering},\ }\href
  {https://doi.org/DOI:10.1063/1.3607438} {\bibfield  {journal} {\bibinfo
  {journal} {Rev. Sci. Instrum.}\ }\textbf {\bibinfo {volume} {82}},\ \bibinfo
  {pages} {073104} (\bibinfo {year} {2011})}\BibitemShut {NoStop}%
\bibitem [{\citenamefont {Achkar}\ \emph {et~al.}(2011)\citenamefont {Achkar},
  \citenamefont {Regier}, \citenamefont {Wadati}, \citenamefont {Kim},
  \citenamefont {Zhang},\ and\ \citenamefont {Hawthorn}}]{Achkar11}%
  \BibitemOpen
  \bibfield  {author} {\bibinfo {author} {\bibfnamefont {A.~J.}\ \bibnamefont
  {Achkar}}, \bibinfo {author} {\bibfnamefont {T.~Z.}\ \bibnamefont {Regier}},
  \bibinfo {author} {\bibfnamefont {H.}~\bibnamefont {Wadati}}, \bibinfo
  {author} {\bibfnamefont {Y.-J.}\ \bibnamefont {Kim}}, \bibinfo {author}
  {\bibfnamefont {H.}~\bibnamefont {Zhang}},\ and\ \bibinfo {author}
  {\bibfnamefont {D.~G.}\ \bibnamefont {Hawthorn}},\ }\bibfield  {title}
  {\bibinfo {title} {Bulk sensitive x-ray absorption spectroscopy free of
  self-absorption effects},\ }\href
  {https://doi.org/10.1103/PhysRevB.83.081106} {\bibfield  {journal} {\bibinfo
  {journal} {Phys. Rev. B}\ }\textbf {\bibinfo {volume} {83}},\ \bibinfo
  {pages} {081106(R)} (\bibinfo {year} {2011})}\BibitemShut {NoStop}%
\bibitem [{\citenamefont {Achkar}\ \emph {et~al.}(2013)\citenamefont {Achkar},
  \citenamefont {He}, \citenamefont {Sutarto}, \citenamefont {Geck},
  \citenamefont {Zhang}, \citenamefont {Kim},\ and\ \citenamefont
  {Hawthorn}}]{Achkar13}%
  \BibitemOpen
  \bibfield  {author} {\bibinfo {author} {\bibfnamefont {A.~J.}\ \bibnamefont
  {Achkar}}, \bibinfo {author} {\bibfnamefont {F.}~\bibnamefont {He}}, \bibinfo
  {author} {\bibfnamefont {R.}~\bibnamefont {Sutarto}}, \bibinfo {author}
  {\bibfnamefont {J.}~\bibnamefont {Geck}}, \bibinfo {author} {\bibfnamefont
  {H.}~\bibnamefont {Zhang}}, \bibinfo {author} {\bibfnamefont {Y.-J.}\
  \bibnamefont {Kim}},\ and\ \bibinfo {author} {\bibfnamefont {D.~G.}\
  \bibnamefont {Hawthorn}},\ }\bibfield  {title} {\bibinfo {title} {Resonant
  x-ray scattering measurements of a spatial modulation of the {Cu $3d$} and {O
  $2p$} energies in stripe-ordered cuprate superconductors},\ }\href
  {https://doi.org/10.1103/PhysRevLett.110.017001} {\bibfield  {journal}
  {\bibinfo  {journal} {Phys. Rev. Lett.}\ }\textbf {\bibinfo {volume} {110}},\
  \bibinfo {pages} {017001} (\bibinfo {year} {2013})}\BibitemShut {NoStop}%
\bibitem [{\citenamefont {Sarrao}\ \emph {et~al.}(1994)\citenamefont {Sarrao},
  \citenamefont {Mandrus}, \citenamefont {Migliori}, \citenamefont {Fisk},
  \citenamefont {Tanaka}, \citenamefont {Kojima}, \citenamefont {Canfield},\
  and\ \citenamefont {Kodali}}]{Sarrao1994}%
  \BibitemOpen
  \bibfield  {author} {\bibinfo {author} {\bibfnamefont {J.~L.}\ \bibnamefont
  {Sarrao}}, \bibinfo {author} {\bibfnamefont {D.}~\bibnamefont {Mandrus}},
  \bibinfo {author} {\bibfnamefont {A.}~\bibnamefont {Migliori}}, \bibinfo
  {author} {\bibfnamefont {Z.}~\bibnamefont {Fisk}}, \bibinfo {author}
  {\bibfnamefont {I.}~\bibnamefont {Tanaka}}, \bibinfo {author} {\bibfnamefont
  {H.}~\bibnamefont {Kojima}}, \bibinfo {author} {\bibfnamefont {P.~C.}\
  \bibnamefont {Canfield}},\ and\ \bibinfo {author} {\bibfnamefont {P.~D.}\
  \bibnamefont {Kodali}},\ }\bibfield  {title} {\bibinfo {title} {Complete
  elastic moduli of
  {${\mathrm{La}}_{2\mathrm{\ensuremath{-}}\mathit{x}}$${\mathrm{Sr}}_{\mathit{x}}$${\mathrm{CuO}}_{4}$}
  (x=0.00 and 0.14) near the tetragonal-orthorhombic structural phase
  transition},\ }\href {https://doi.org/10.1103/PhysRevB.50.13125} {\bibfield
  {journal} {\bibinfo  {journal} {Phys. Rev. B}\ }\textbf {\bibinfo {volume}
  {50}},\ \bibinfo {pages} {13125} (\bibinfo {year} {1994})}\BibitemShut
  {NoStop}%
\bibitem [{\citenamefont {Eisebitt}\ \emph {et~al.}(1993)\citenamefont
  {Eisebitt}, \citenamefont {B\"oske}, \citenamefont {Rubensson},\ and\
  \citenamefont {Eberhardt}}]{Eisebitt93}%
  \BibitemOpen
  \bibfield  {author} {\bibinfo {author} {\bibfnamefont {S.}~\bibnamefont
  {Eisebitt}}, \bibinfo {author} {\bibfnamefont {T.}~\bibnamefont {B\"oske}},
  \bibinfo {author} {\bibfnamefont {J.-E.}\ \bibnamefont {Rubensson}},\ and\
  \bibinfo {author} {\bibfnamefont {W.}~\bibnamefont {Eberhardt}},\ \bibfield  {title} {\bibinfo {title} {Determination of absorption coefficients for concentrated samples by fluorescence detection},\ }}\href
  {https://doi.org/10.1103/PhysRevB.47.14103} {\bibfield  {journal} {\bibinfo
  {journal} {Phys. Rev. B}\ }\textbf {\bibinfo {volume} {47}},\ \bibinfo
  {pages} {14103} (\bibinfo {year} {1993})}\BibitemShut {NoStop}%
\bibitem [{\citenamefont {Chantler}(2000)}]{Chantler00}%
  \BibitemOpen
  \bibfield  {author} {\bibinfo {author} {\bibfnamefont {C.}~\bibnamefont
  {Chantler}},\ }\bibfield  {title} {\bibinfo {title} {Detailed tabulation of
  atomic form factors, photoelectric absorption and scattering cross section,
  and mass attenuation coefficients in the vicinity of absorption edges in the
  soft x-ray($z$=30-36, $z$=60-89, $e$=0.1 kev-10 kev), addressing convergence
  issues of earlier work.},\ }\href{https://doi.org/10.1063/1.1321055} {\bibfield  {journal} {\bibinfo
  {journal} {J. Phys. Chem. Ref. Data}\ }\textbf {\bibinfo {volume} {29}},\
  \bibinfo {pages} {597} (\bibinfo {year} {2000})}\BibitemShut {NoStop}%
\end{thebibliography}
\end{document}


\title{Supplementary Information: Tuning charge density wave order and structure via uniaxial stress in a stripe-ordered cuprate superconductor}

\author{Naman~K.~Gupta}
\affiliation{Department of Physics and Astronomy, University of Waterloo, Waterloo, Ontario N2L 3G1, Canada}

\author{Ronny~Sutarto}
\affiliation{Canadian Light Source, University of Saskatchewan, Saskatoon, Saskatchewan, S7N 2V3, Canada}

\author{Rantong~Gong}
\affiliation{Department of Physics and Astronomy, University of Waterloo, Waterloo, Ontario N2L 3G1, Canada}

\author{Stefan~Idziak}
\affiliation{Department of Physics and Astronomy, University of Waterloo, Waterloo, Ontario N2L 3G1, Canada}

\author{Hiruy~Hale}
\affiliation{Department of Physics and Astronomy, University of Waterloo, Waterloo, Ontario N2L 3G1, Canada}

\author{Young-June~Kim}
\affiliation{Department of Physics, University of Toronto, Toronto, M5S 1A7, Canada}

\author{David~G. Hawthorn}
\affiliation{Department of Physics and Astronomy, University of Waterloo, Waterloo, Ontario N2L 3G1, Canada}

\maketitle

\section{Methods}

Resonant x-ray scattering measurements were performed at the REIXS beamline at the Canadian Light Source \cite{Hawthorn11a}, on a sample of La$_{1.475}$Nd$_{0.4}$Sr$_{0.125}$CuO$_4$, fragments of which have been studied in past reports of x-ray absorption spectroscopy~\cite{Achkar11} and resonant x-ray scattering~\cite{Achkar13,Achkar16}. The CDW (-0.24 0 1.5) and (0 -0.24 1.5) Bragg peaks were investigated at a photon energy corresponding to the peak of the Cu-$L_3$ absorption edge (931.3 eV). Note, measurements of the CDW peaks are shown as raw data, normalized only to the incident beam intensity, but otherwise free of any background subtraction. In addition, to examine the LTO to LTT structural phase transition, we measured the (0~0~1) Bragg peak beak at the Cu-$L$ edge (931.3 eV) and at an energy associated with the apical oxygen (533.3 eV). The sample was orientated by measuring the (0 0 4), (-1 0 3) and (0 -1 3) structural Bragg peaks at 2200 eV, with all Bragg peaks indexed to the high temperature tetragonal phase.  

The sample was cut and polished to form a rectangular $ab$ face sample of dimensions 2 x 0.5 x 0.2 mm. Unixial stress was applied parallel to the longest axis of the sample, which is oriented parallel to [100]. Subsequent to polishing, the sample's surface was etched with bromine. The application of uniaxial stress was applied by mounting the sample on a custom mechanically-actuated stage, depicted in Figure S1. The sample is mounted on a BeCu plate that has been machined with notch flexures to enable motion along a single axis, but provide a degree of stiffness to bending or tilting. The sample is positioned in a groove in the BeCu plate, with the edges encapsulated above and below with EPO-TEK H2OE silver epoxy in an effort to apply uniform stress. The BeCu plate is affixed to two Cu blocks that can be translated relative to one another by actuating a differential screw that enables small controlled displacements between the plates, resulting in uniaxial compressive or tensile stress on the sample. A novel feature of the strain device is the capability to rotate the device azimuthally under stress, in-vacuum and at low temperature in order to orient the direction of the applied stress either parallel or perpendicular to the scattering plane of the  diffractometer. As such, stress can be applied to the sample along the $a$-axis and CDW peaks can be investigated along both \mbox{(-0.24~0~1.5)} and \mbox{(0~-0.24~1.5)} without needing to thermally cycle or remount the sample.  

Unfortunately, our device does not measure the magnitude of the applied stress. However, by measuring the structural Bragg peaks before and after the application of uniaxial stress, we can identify that the peak positions do not shift by amounts larger than the accuracy of our measurements. This limits the in-plane (or $c$-axis) compressive strain to be less than 0.2$\%$. Given the $C_{11} = 267$ GPa in related materials~\cite{Sarrao1994}, this strain limit corresponds to an uniaxial stress along [100] of $< 0.5$ GPa.

    \begin{figure}[ht]
     \begin{center}
   \includegraphics[width=\columnwidth]{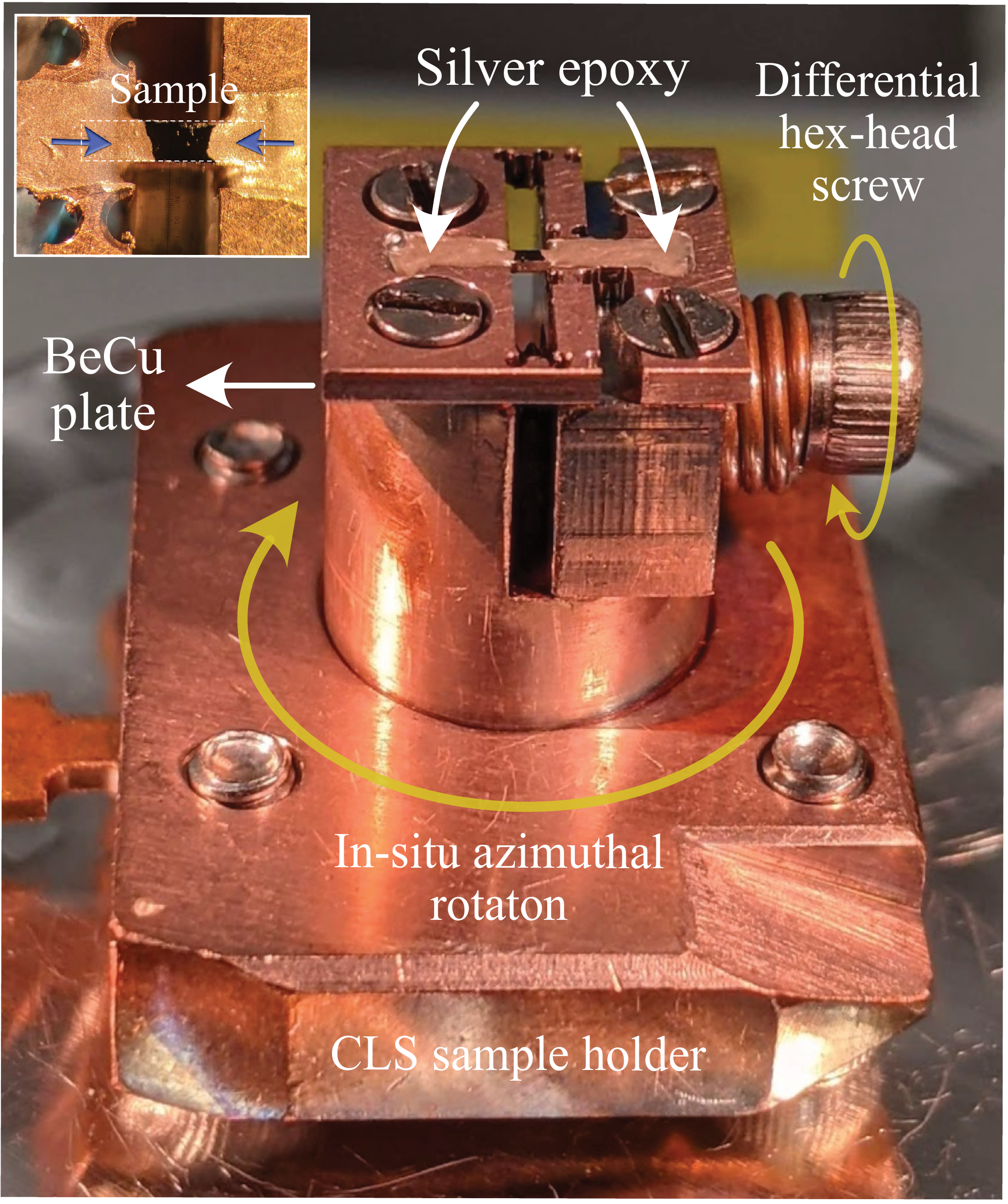}
        \caption{Schematic of the uniaxial stress device. Rotation of a differential screw is used to apply uniaxial stress to a sample that is mounted to a BeCu plate. The device can be actuated in vacuum and at low temperatures in order to change the applied stress or to rotate the device azimuthally to orient the strain parallel or perpendicular to the horizontal scattering plane of the diffractometer. The sample shown in the inset is set in position using silver epoxy.
        \label{fig:figS1}}
        \end{center}
    \end{figure}

\section{Fluorescent background}
The measurements of the CDW peaks depicted in figure 3 of the main text show a CDW peak at low temperatures that is unresolved at 90 K. In addition to more well-resolved CDW peaks at low temperatures, several studies have also reported the observation of low intensity, broad in $Q$ features that have been interpreted as evidence for charge density fluctuations or a weak, short-range from of CDW order~\cite{Arpaia21,Boyle2021}. In our measurements, we are unable to unambiguously identify evidence for charge density fluctuations above the nominal $T_{CDW}$ in our sample.

\begin{figure}[ht]
     \begin{center}
   \includegraphics[width=\columnwidth]{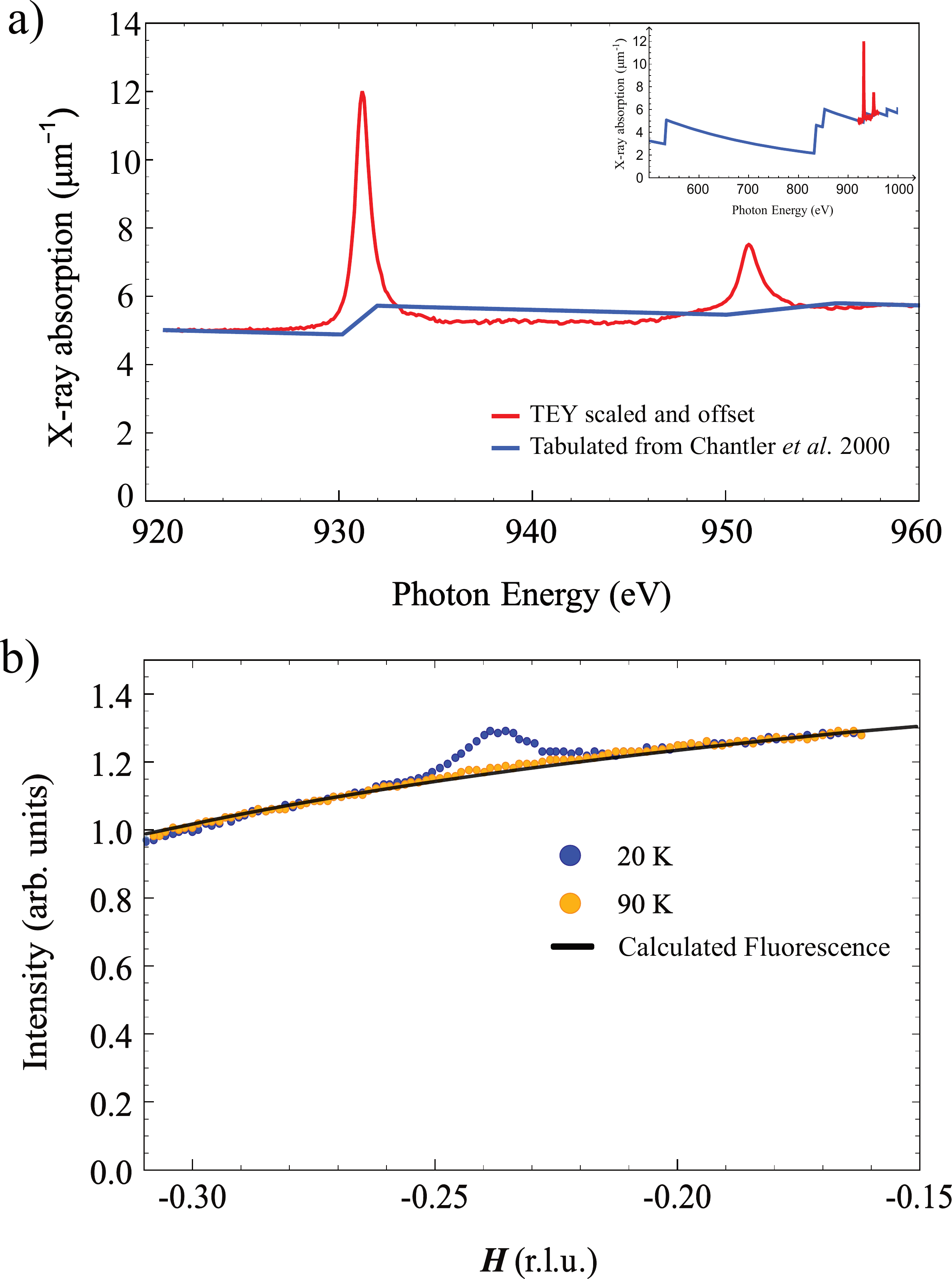}
        \caption{a) Linear x-ray absorption co-efficient, $\mu(E)$, of LNSCO as a function of photon energy from tabulated values~\cite{Chantler00} and TEY measurements of a freshly cleaved sample.  The TEY data is scaled and offset to match the magnitude of tabulated $\mu(E)$ above and below the Cu $L$ edges.  b) $H$ dependence of the measured intensity at 20 K and 90 K for unstrained sample (UNS2). The scan was a $\theta$ scan at a fixed value of $2\theta = 119.9^\circ$.  The calculated fluorescence from eqn.~\ref{eqn:eqS1} agrees well with the measured intensity at 90 K, with residuals less than 1\% of the total intensity.
        \label{fig:figS2}}
        \end{center}
    \end{figure}

The measured intensity at 90 K is to be dominated by x-ray fluorescence, as shown in figure~\ref{fig:figS2}.  For $\sigma$ polarized light, the expected total fluorescent background can be calculated using the formula \cite{Eisebitt93,Achkar11}:
\begin{equation}
	 \frac{I}{I_0} \!=\! C \sum_{X} \frac{A_X}{1+\frac{\mu(E_f)}{\mu(E_f)}\frac{\sin \alpha}{\sin \beta}}.
\label{eqn:eqS1}
\end{equation}
where $\mu(E_i)$ and $\mu(E_f)$ are the total linear attenuation coefficients of the material at the incident and emitted photon energies, $E_i$ and $E_f$, and $\alpha$ and $\beta$ are the angle of incidence and angle of detection, respectively~\cite{Eisebitt93,Achkar11}. The intensity is summed over the emission from all elements in the material and all core electrons of those elements, denoted by $X$ ($X$ = Cu $2p$, Cu $3p$, O $1s$, O $2s$, ...), with the relative intensity of the emission from element/core electron $X$ parameterized by $A_X$.

Apart from an overall scaling factor, the parameters needed to calculate the fluorescence from eqn.~\ref{eqn:eqS1} can determined from the experiment. To determine $A_X$,  previously measured x-ray emission in LNSCO for incident photon energy at the peak of the Cu $L_3$ edge~\cite{Achkar11}, identified x-ray emission peaks at the following energies: O $K_\alpha = 525$ eV, La $M_\zeta = 630$ eV, La  $M_{\alpha, \beta} \simeq 820$ eV and Cu $L_{\alpha, \beta} \simeq 925$ eV. A multi-peak fit to the intensity of these x-ray emission lines gives $A_{Cu_{L_{\alpha, \beta}}} = 1$, this fit gives $A_{O K}=0.073$, $A_{LaM_\zeta} = 0.032$ and $A_{LaM_{\alpha,\beta}}= 0.128$.

To estimate the values of $\mu(E_i)$ and $\mu(E_i)$, a combination of TEY measurement of the XAS on a freshly cleaved sample, as well as tabulated values of the absorption co-efficient from NIST are used~\cite{Chantler00}. To determine the absorption coefficient at the peak of the Cu $L_3$ edge,  $\mu(931.3 eV)$, the TEY signal is scaled and offset to match the tabulated values of the absorption co-efficient from NIST.  For the absorption co-efficents at the emission energies, the tabulated values from NIST are utilized.  This gives $\mu(931.3 eV) = 12 ~\mathrm{\mu m}^{-1}$, $\mu(525 eV) = 3.03 ~\mathrm{\mu m}^{-1}$, $\mu(820 eV) = 2.21 ~\mathrm{\mu m}^{-1}$, $\mu(630 eV) = 3.77 ~\mathrm{\mu m}^{-1}$ and  $\mu(925 eV) = 5.01 ~\mathrm{\mu m}^{-1}$.

Using these values, the fluorescence as a function of $H$ is calculated using eq.~\ref{eqn:eqS1} and scaled to match the overall intensity of the measured single at 90 K (a measurement along $H$ for the sample in the unstained state was chosen for this purpose), as shown in figure~\ref{fig:figS2}.  At 90 K, the measurement and calculation agree well, with residuals of order 1\% of the total intensity, indicating that fluorescence dominates the measured signal. The variance between the calculation and measurement is comparable to the $\sim 1\%$ variation in background between scans observed for this series of measurements, as shown in figure~3 of the main text.

Other factors that also contribute to the measured intensity and are not included in the model include emission from excitations (phonons, plasmons, paramagnons) and diffuse scattering from crystalline defects or thermal diffuse scattering.  Moreover, extrinsic factors that may produce angle-dependent variations that are not included in the model include instability of the beam position or energy, tails of the beamspot exceeding the sample dimensions or tails of the incident or emitted beam being shadowed by the silver epoxy utilized to embed the sample. Consequently, additional scattering that may result from charge density fluctuations is less than or equal to the estimated accuracy of the measurements and cannot be unambiguously resolved in the present measurements.